\documentclass[10pt,journal,compsoc,twocolumn]{IEEEtran}
\IEEEoverridecommandlockouts
\ifCLASSOPTIONcompsoc
  \usepackage[nocompress]{cite}
\else
  \usepackage{cite}
\fi
\ifCLASSINFOpdf
\else
\fi
\usepackage{balance}
\usepackage[absolute]{textpos}
\usepackage{tcolorbox}
\usepackage{graphicx}
\usepackage{amsmath}
\usepackage{amssymb}
\usepackage{color}
\usepackage{times}
\usepackage{caption}
\usepackage{booktabs}
\usepackage{multirow}
\usepackage{footmisc}

\usepackage{lipsum,microtype}
\DeclareUnicodeCharacter{0394}{$\Delta$}
\DeclareUnicodeCharacter{03A3}{$\Sigma$}

\DeclareRobustCommand*{\IEEEauthorrefmark}[1]{\raisebox{0pt}[0pt][0pt]{\textsuperscript{\footnotesize\ensuremath{#1}}}}

\begin{document}

\title{The Dawn of AI-Native EDA: Opportunities and Challenges of Large Circuit Models}

\author{
\IEEEauthorblockN{Tsung-Yi Ho\IEEEauthorrefmark{1}, Sadaf Khan\IEEEauthorrefmark{1}, Jinwei Liu\IEEEauthorrefmark{1}, Yu Li\IEEEauthorrefmark{1}, Yi Liu\IEEEauthorrefmark{1}, Zhengyuan Shi\IEEEauthorrefmark{1}, Ziyi Wang\IEEEauthorrefmark{1}, Qiang Xu\IEEEauthorrefmark{1}\textsuperscript{\textsection}, \\ Evangeline F.Y. Young\IEEEauthorrefmark{1}, Bei Yu\IEEEauthorrefmark{1}, Ziyang Zheng\IEEEauthorrefmark{1}, Binwu Zhu\IEEEauthorrefmark{1}, Keren Zhu\IEEEauthorrefmark{1}}
\IEEEauthorblockA{\\ \IEEEauthorrefmark{1}\textbf{The Chinese University of Hong Kong}
}
\vspace{12pt}\\
\IEEEauthorblockN{Yiqi Chen\IEEEauthorrefmark{2}, Ru Huang\IEEEauthorrefmark{2,3}, Yun Liang\IEEEauthorrefmark{2}, Yibo Lin\IEEEauthorrefmark{2}, Guojie Luo\IEEEauthorrefmark{2}\textsuperscript{\textsection}, Guangyu Sun\IEEEauthorrefmark{2}, Runsheng Wang\IEEEauthorrefmark{2}, Xinming Wei\IEEEauthorrefmark{2}, Chenhao Xue\IEEEauthorrefmark{2}, Jun Yang\IEEEauthorrefmark{3}, Haoyi Zhang\IEEEauthorrefmark{2}, Zuodong Zhang\IEEEauthorrefmark{2}, Yuxiang Zhao\IEEEauthorrefmark{2}, Sunan Zou\IEEEauthorrefmark{2}}
\IEEEauthorblockA{\\ \IEEEauthorrefmark{2}\textbf{Peking University} \IEEEauthorrefmark{3}\textbf{Southeast University}
}
\vspace{12pt}\\
\IEEEauthorblockN{Lei Chen\IEEEauthorrefmark{4}, Yu Huang\IEEEauthorrefmark{5}, Min Li\IEEEauthorrefmark{4}, Dimitrios Tsaras\IEEEauthorrefmark{4}, Mingxuan Yuan\IEEEauthorrefmark{4}\textsuperscript{\textsection}, Hui-Ling Zhen\IEEEauthorrefmark{4}}
\IEEEauthorblockA{\\ \IEEEauthorrefmark{ }\IEEEauthorrefmark{4}\textbf{Huawei Noah's Ark Lab} \IEEEauthorrefmark{5}\textbf{Huawei HiSilicon}
}
\vspace{12pt}\\
\IEEEauthorblockN{Zhufei Chu\IEEEauthorrefmark{6}, Wenji Fang\IEEEauthorrefmark{7}, Xingquan Li\IEEEauthorrefmark{8}, Junchi Yan\IEEEauthorrefmark{9}, Zhiyao Xie\IEEEauthorrefmark{7}}, Xuan Zeng\IEEEauthorrefmark{10}\\
\IEEEauthorblockA{\IEEEauthorrefmark{6}\textbf{Ningbo University} \IEEEauthorrefmark{7}\textbf{Hong Kong University of Science and Technology} \\
\IEEEauthorrefmark{8}\textbf{Peng Cheng Laboratory}
\IEEEauthorrefmark{9}\textbf{Shanghai Jiao Tong University}
\IEEEauthorrefmark{10}\textbf{Fudan University}}

}

\IEEEtitleabstractindextext{%
\begin{center}
    \setcounter{figure}{0}
    \centering
    \includegraphics[width=\linewidth]{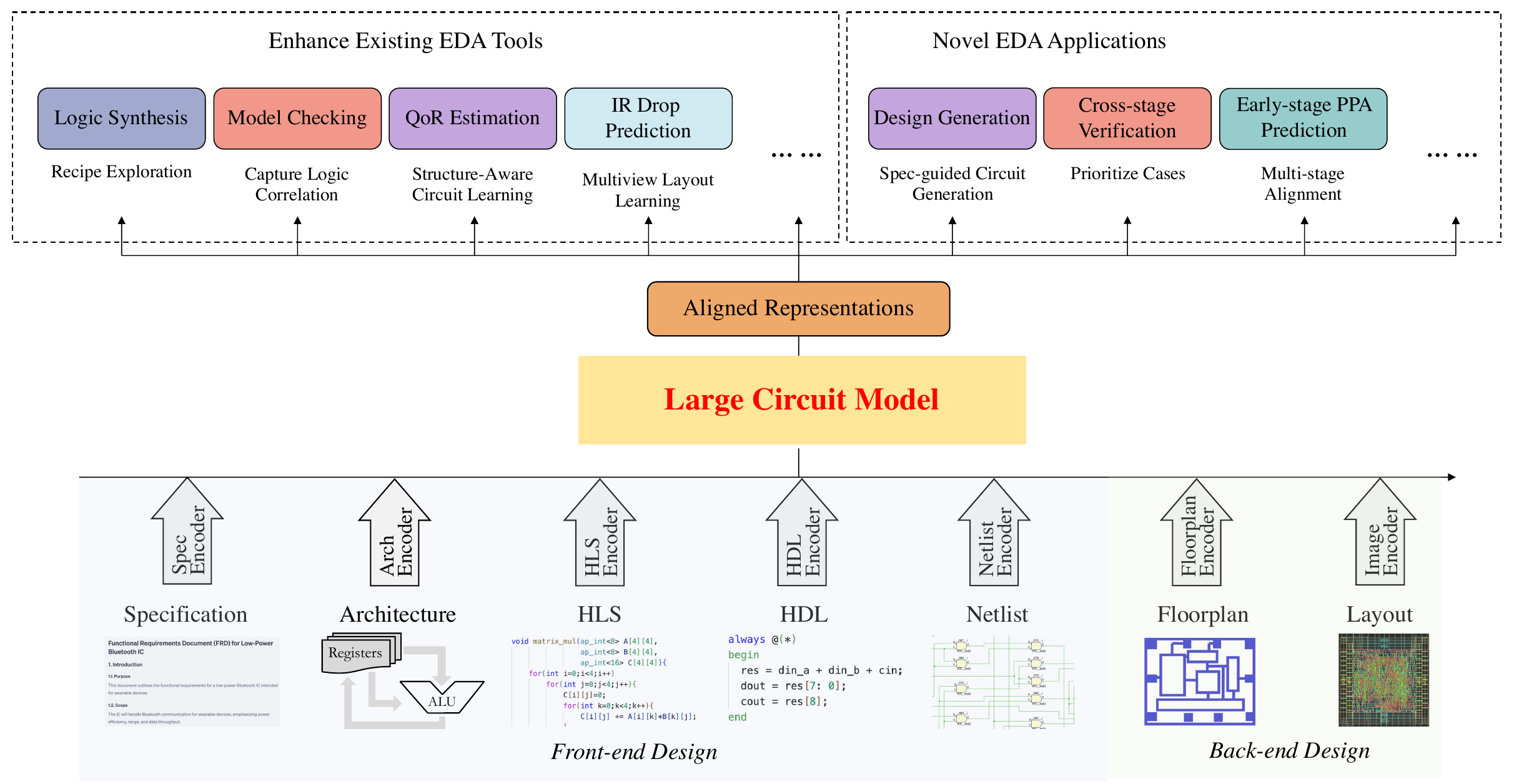}
    \captionsetup{type=figure, margin=1pt}
    \captionof{figure}{\textbf{Large Circuit Models}: we call upon the creation of dedicated foundation models for circuits, which intricately intertwines computation with structure, unlike other types of data (e.g., texts and images). Specifically, each design stage of the EDA flow is considered a separate modality and requires a specific representation learning strategy to embed the available circuit characteristics. The higher the design level is, the more semantics to represent; The lower the design level is, the more details to represent. Central to the appeal of LCMs is their ability to fuse and align disparate representations throughout the design continuum, creating a unified narrative that spans from high-level functional specifications to detailed physical layouts. This unified approach promises to streamline the EDA process, reduce time-to-market, and improve design PPA.      
 }
    \label{fig:overview}
\end{center}
}

\begin{textblock*}{\textwidth}(1.5cm, \textheight)
\begin{tcolorbox}[width=\textwidth,nobeforeafter,colback=white,boxrule=0pt,colframe=white]
{\textsection}\footnotesize{\ The authors in each institution are ordered alphabetically. \textbf{Contact}: qxu@cse.cuhk.edu.hk; gluo@pku.edu.cn; yuan.mingxuan@huawei.com.}
\end{tcolorbox}
\end{textblock*}

\maketitle

\twocolumn[{

\begin{abstract}

  Within the Electronic Design Automation (EDA) domain, AI-driven solutions have emerged as formidable tools, yet they typically augment rather than redefine existing methodologies. These solutions often repurpose deep learning models from other domains such as vision, text, and graph analytics applying them to circuit design without tailoring to the unique complexities of electronic circuits. Such an ``AI4EDA'' approach falls short of achieving a holistic design synthesis and understanding, overlooking the intricate interplay of electrical, logical, and physical facets of circuit data. \textbf{This paper argues for a paradigm shift from AI4EDA towards AI-native EDA}, integrating AI at the core of the design process. Pivotal to this vision is the development of a multimodal circuit representation learning technique, poised to provide a comprehensive understanding by harmonizing and extracting insights from varied data sources, such as functional specifications, RTL designs, circuit netlists, and physical layouts. \\

  We champion the creation of \textbf{large circuit models (LCMs)} that are inherently multimodal, crafted to decode and express the rich semantics and structures of circuit data, thus fostering more resilient, efficient, and inventive design methodologies. Embracing this AI-native philosophy, we foresee a trajectory that transcends the current innovation plateau in EDA, igniting a profound ``shift-left'' in electronic design methodology. The envisioned advancements herald not just an evolution of existing EDA tools but a revolution, giving rise to novel instruments of design tools that promise to radically enhance design productivity and inaugurate a new epoch where the optimization of circuit performance, power, and area (PPA) is achieved not incrementally, but through leaps that redefine the benchmarks of electronic systems’ capabilities.
\end{abstract}
\hfill
\hrule
\hfill
\\
}]

\ifCLASSOPTIONcaptionsoff
  \newpage
\fi


\section{The Foundation Model Paradigm}\label{sec:Introduction}

The landscape of artificial intelligence (AI) has been profoundly transformed in recent years by the advent of large foundation models. These models, characterized by their vast scale and general applicability, have demonstrated an uncanny ability to understand, predict, and generate content with a level of sophistication that was previously the exclusive domain of human intelligence.

\subsection{The Rise of Foundation Models}

Large foundation models represent a significant leap in AI. These models, typically pre-trained on web-scale datasets using self-supervision techniques~\cite{bommasani2021opportunities}, have been adapted to excel in a wide array of downstream tasks. In the fields of natural language processing (NLP) and computer vision (CV), these models have not only set new benchmarks but have fundamentally redefined the realms of possibility.

In NLP, models like BERT~\cite{devlin-etal-2019-bert} and its derivatives, including RoBERTa~\cite{liu2019roberta} and T5~\cite{raffel2020exploring}, have revolutionized language understanding, especially in contextual interpretation of text, thereby enhancing complex language-based tasks. Concurrently, the decoder-only GPT series~\cite{brown2020language} has shown remarkable versatility, excelling in diverse tasks from creative writing to code generation and pointing towards the burgeoning potential of artificial general intelligence (AGI). In CV area, self-supervised foundation models~\cite{chen2020simple,he2020momentum,he2022masked} have achieved competitive performances in image understanding tasks, rivaling fully supervised approaches.

The recent advent of multimodal foundation models has ushered in a new era of possibilities, integrating diverse data types such as text, images, and audio. A pioneering example is the CLIP model~\cite{radford2021learning}, which effectively bridges linguistic and visual data through contrastive learning. This innovation has set the stage for generative models like DALL-E~\cite{ramesh2021zero} and Stable Diffusion~\cite{rombach2022high}, which demonstrate the capability to generate intricate images from textual descriptions, seamlessly blending visual and linguistic understanding. Additionally, the recently introduced promptable CV systems (e.g., SAM~\cite{kirillov2023segment}) have exhibited exceptional zero-shot generalization in image segmentation, enabling precise object identification and extraction. The emergence of GPT-4V~\cite{yang2023dawn} and Gemini~\cite{team2023gemini} further exemplify the evolution of AI, seamlessly navigating and synthesizing multimodal information, thereby opening new avenues for innovation across various fields, from creative content generation to complex problem-solving in engineering and design.

Despite these advancements, the field of circuit design has only begun to scratch the surface of what foundation models can offer. This hesitant engagement contrasts starkly with the transformative potential these models hold for this important field.

\subsection{The Unique Challenge of Circuit Data}

In the realm of circuit design, a notable phenomenon is the inherent similarity of many new designs to past iterations. Despite these similarities, designers frequently face the challenge of recreating or redesigning circuits from scratch, driven by the subtle yet critical nuances required to meet ambitious performance, power, and area (PPA) objectives. This repetitive process highlights the need for a learning solution that can effectively draw from historical successes and failures.

The emergence of AI for electronic design automation (AI4EDA) solutions~\cite{rapp2021mlcad} marks an attempt to integrate machine learning (ML) techniques into circuit design. These advancements represent significant progress but often only augment, rather than redefine, existing methodologies. Typically, AI4EDA repurposes deep learning models from other
domains for EDA tasks such as PPA estimation and optimization, verification, or fault detection. However, within the confines of traditional design frameworks, these models act more as individual analytical tools than as integral components of the design process, often failing to fully address the unique complexities of circuit data.

Specifically, the distinctive nature of circuit data poses unique challenges for machine learning. Unlike text, images, or regular graph data, \textbf{circuit design intricately intertwines computation with structure}. Minor structural changes can lead to significant functional impacts, and vice versa. This interdependency renders the task of modeling circuits highly nuanced and complex. Without considering the above, existing AI4EDA solutions frequently fall short in achieving a comprehensive synthesis and understanding of the multifaceted interplay between electrical, logical, and physical aspects of circuit data, which is essential for truly innovative design synthesis.

Recent advancements in AI-native circuit representation learning, such as those presented in~\cite{li2022deepgate,shi2023deepgate2}, have begun to address these unique challenges. The integration of multimodal learning presents a significant opportunity to further enhance their effectiveness. By adopting the principles and capabilities demonstrated by existing foundation models on various types of data, we conceptualize \textbf{a paradigm shift from AI4EDA to AI-native EDA}. 

Pivotal to this vision is the development of sophisticated \textbf{large circuit models (LCMs)}. Envisioned as models adept at integrating and interpreting diverse data types specific to circuit design, LCMs could potentially revolutionize the design, optimization, and verification processes of electronic circuits.

\subsection{The Feasibility and Promises of AI-Native LCMs}

In the world of semiconductor design, the potential for leveraging large circuit models is not just aspirational; it is rooted in a rich heritage of technological evolution. 

Decades of research and development have yielded a vast repository of circuit data. Though proprietary barriers exist, there is enough in the public domain~\cite{OpenCores,XiangShan,RocketChip} to fuel the development of robust, intelligent models. The industry's long history provides data that is richly annotated with domain expertise, offering deep insights into the intricacies of circuit design.

Moreover, the landscape of circuit types, though vast, is marked by commonalities that transcend individual designs. Processors, domain accelerators (e.g. digital signal processors (DSPs) and AI accelerators), communication modules, and other core components display a pattern of design module reuse. Examples of these reusable modules include arithmetic units, various decoders, and cryptographic cores. This consistency provides a predictable pattern—akin to an inductive bias—that is conducive to the application of machine learning models.

Advances in neural network architectures, particularly Transformers~\cite{vaswani2017attention} and graph neural networks (GNNs)~\cite{zhou2020graph}, are well-suited to capturing the complex, graph-like structure of circuit schematics. They present an opportunity to transform the intricate web of design elements into actionable insights, a feat previously unattainable. The AI advancements from other domains, e.g., CLIP model with multimodal machine learning capabilities~\cite{baltruvsaitis2018multimodal} and large language models for code generation~\cite{austin2021program}, further underscore the potential for transformative applications in LCMs. These capabilities could be adapted to address the unique challenges in circuit designs of various forms, enabling more nuanced and comprehensive modeling than ever before.

In summary, while the challenges are nontrivial, the development of LCMs is poised on a solid foundation of historical data, pattern prevalence, and cutting-edge computational techniques. The potential for LCMs to revolutionize the field of EDA is not just a theoretical possibility but a tangible goal, driven by the convergence of historical knowledge and modern AI advancements. By processing and interpreting a diverse array of data sources and formats, including schematic diagrams, textual specifications, register-transfer level (RTL) designs, circuit netlists, physical layouts, and performance metrics, LCMs can facilitate a `shift-left' in the design methodology. This proactive AI-native approach enables the early identification of potential performance issues and design bottlenecks, streamlining the testing and redesign processes, and leading to more informed and efficient development cycles.

\subsection{Overview of This Perspective Paper}

This paper embarks on a comprehensive exploration into the dawn of AI-native EDA, focusing on the development and application of large circuit models that inherently incorporate multimodal data. Spanning nine sections, the paper delves into the historical evolution of EDA, the current state of AI in this field, and the promising future shaped by LCMs.

Section~\ref{sec:History} provides a historical overview of EDA, tracing its evolution alongside the semiconductor industry. It emphasizes how the field has navigated challenges of complexity through abstractions, setting a foundation for understanding the significance of LCMs in this evolving landscape. Next, we discuss the current integration of AI in EDA in Section~\ref{sec:AI4EDA}, highlighting how deep learning has been utilized to improve EDA processes. 

In Section~\ref{sec:LCM}, we introduce AI-native LCMs, illustrating their departure from traditional AI4EDA approaches. It delves into how these models encapsulate the intricacies of circuit design, offering a more comprehensive approach to circuit analysis and even creation. Focusing on the development of unimodal circuit representation learning, Section~\ref{sec:UniLearning} discusses its critical role in building the foundation for multimodal LCMs. It explores the nuances of this approach in achieving a thorough understanding of circuit data. Then, Section~\ref{sec:Multimodality} navigates the transition to multimodal integration in LCMs. It discusses the development of techniques to align and integrate representations from different design stages, emphasizing the importance of preserving the original design intent.

Section~\ref{sec:Cases} illustrates the potential applications of LCMs through case studies and envisioned scenarios, bridging the gap between theoretical concepts and practical implementations. 
In Section~\ref{sec:Special}, we explore the application of LCMs in specialized circuit domains, discussing how these models can be adapted to cater to the unique needs of diverse circuit types other than standard digital circuits, including standard cell designs, datapath units, and analog circuits.

Next, we discuss the challenges and opportunities presented by the adoption of LCMs in EDA in Section~\ref{sec:Challenges}. It highlights issues such as data scarcity and scalability, as well as the potential advancements these challenges can foster. Finally, the paper concludes with a summary of the key insights and a forward-looking perspective in Section~\ref{sec:Future}. It calls for continued collaboration between the AI and EDA communities and suggests future research avenues to further advance the field.

\section{Historical Odyssey of EDA}\label{sec:History}

As we stand on the precipice of this new frontier of AI-native EDA, it is vital to appreciate the historical EDA journey. Understanding the evolution of cutting-edge EDA tools, methodologies, and philosophies will provide invaluable context for the challenges and opportunities that lie ahead.

\subsection{Core Objectives and Complexities in EDA}

The odyssey of EDA is a chronicle of human ingenuity and technological advancement. It is a story that mirrors the exponential growth of the semiconductor industry, fueled by Moore's Law, and characterized by the ceaseless push for smaller, faster, and more efficient electronic devices. The journey from simple logic circuits to today's billion-transistor integrated circuits (ICs) has necessitated a layered hierarchical design methodology with the help of sophisticated EDA toolsets. This hierarchy, marked by stages such as specification, architecture design, high-level algorithm design, RTL design, logic synthesis, and physical design, allows for incremental refinement of the circuit design, each stage adding a layer of detail, ensuring functionality while striving for optimization.

The journey of EDA is not just marked by the sophistication of its tools but also by the fundamental goals that drive its evolution. Two core objectives have consistently shaped the development of EDA solutions:

\begin{itemize}
    \item \textbf{Equivalence and Consistency across Transformations}: Ensuring that each transformation—from behavioral descriptions to gate-level implementation and from logical to physical representation—maintains the original design intent is essential.
        C-RTL equivalence checking, assertion-based verification (ABV), logic equivalence checking (LEC), sequential equivalence checking (SEC), and various types of simulation tools have been indispensable in this regard, providing designers with the assurance that despite the myriad of transformations a design undergoes, the end result is functionally equivalent to the original specifications. This integrity across various stages, including architecture design, logic synthesis, technology mapping, and place-and-route, is the bedrock upon which reliable electronic design is built.
    \vspace{5pt}
    \item \textbf{Optimization of PPA and Other Design Factors}: The relentless pursuit of optimizing performance, power, and area is central to EDA. As designs scale and complexities increase, the balance between these three aspects becomes more challenging to achieve.
        Tools dedicated to PPA optimization employ a variety of techniques, including predictive modeling, heuristic algorithms, and iterative refinement, to squeeze out efficiencies at every level of design.
        Meanwhile, the traditional PPA triad is no longer the sole focus.
        With the advent of ultra-deep submicron technologies, new concerns have emerged.
        Circuit reliability has taken center stage, with issues such as electromigration and thermal effects becoming critical. Manufacturability is another growing concern, as variability in fabrication processes can significantly impact yield and performance. 
\end{itemize}

In the fiercely competitive realm of electronic product development, reducing time-to-market (TTM) is paramount. The rapid evolution of consumer electronics, exemplified by the yearly refresh cycles of smartphones and wearables, underscores the urgency to expedite product launches to capture market share and meet consumer expectations. This pressure significantly impacts the EDA process, where the need for TTM can sometimes compromise design thoroughness, leading to potential flaws. For instance, under the gun to release the next generation of microprocessors, teams may bypass exhaustive verification in favor of meeting launch windows, risking the introduction of bugs into the final product. When such issues are not amendable through engineering change orders (ECO)~\cite{lin1995logic}, they necessitate a costly and time-consuming redesign, further exacerbating time-to-market pressures. Therefore, this cycle highlights the crucial need for EDA solutions that not only streamline design and verification processes but also ensure design accuracy from the outset.

\subsection{EDA for Front-End Design} 

In the 1980s, the growth of the semiconductor sector was hindered by the manual creation of large schematics, significantly limiting design productivity~\cite{de2010chip}.
The narrative of front-end EDA tools is a testament to the field's evolution from the era of hand-drawn schematics to the sophistication of automated logic synthesis. This evolution has been underpinned by the introduction of hardware description languages (HDLs) like Verilog and VHDL, which have become the bedrock for digital design representation, simulation, and verification.

\begin{figure*}[!t]
    \centering
    \includegraphics[width=5in]{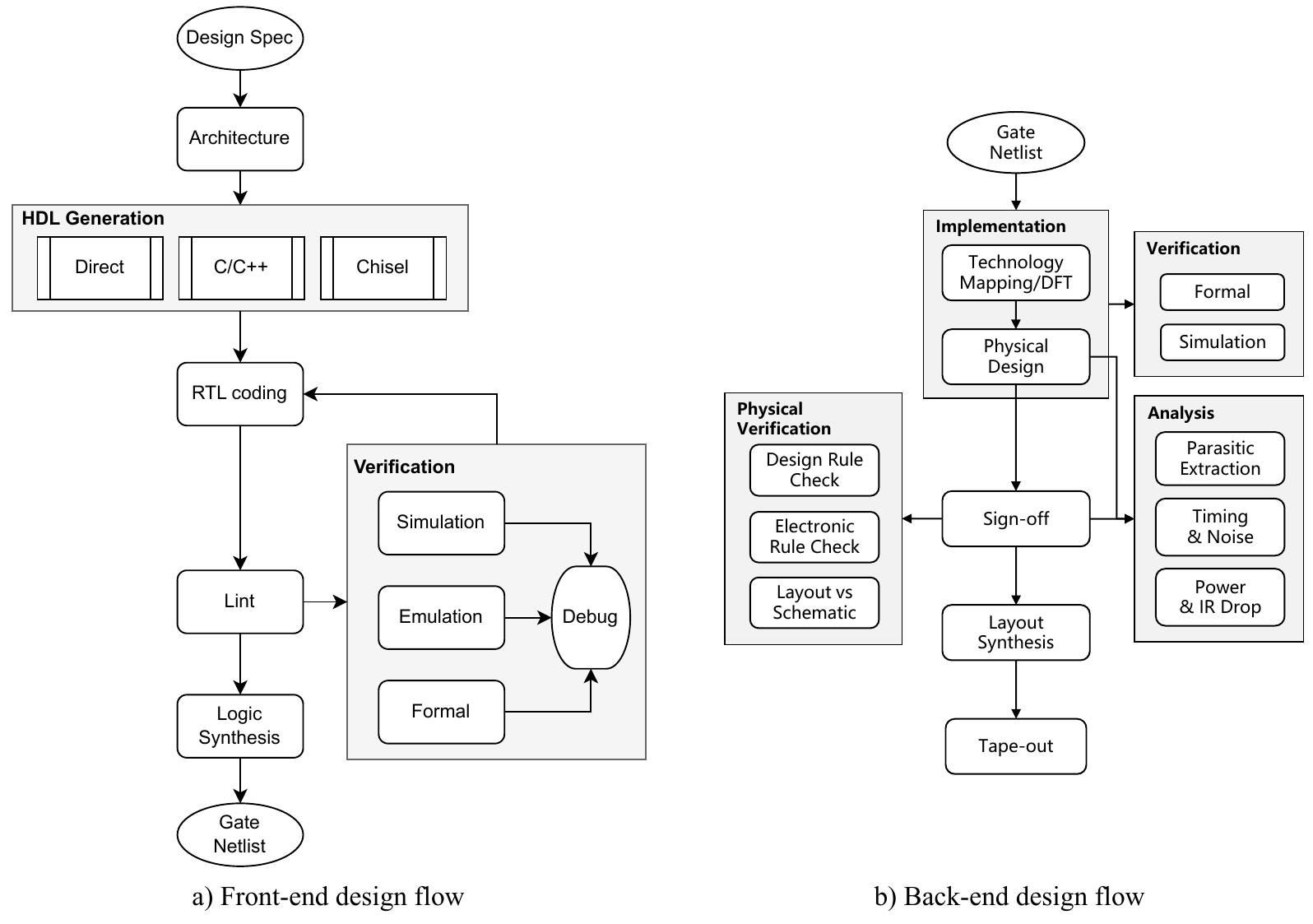}
    \caption{Typical front-end and back-end design flows.}
    \label{fig:fe_be_flow}
\end{figure*}

A typical front-end design flow, also known as logic design, is shown in Fig.~\ref{fig:fe_be_flow} a), in which the design specification is transformed into a logic netlist.
The front-end design flow begins with a design specification, followed by architecture exploration. Subsequently, HDLs are created to translate the design into a form suitable for implementation, typically at the RTL abstraction level. 
The introduction of hardware construction languages (like Chisel~\cite{Chisel}) and C/C++ high-level synthesis (HLS) adds a new dimension to front-end design and offers more flexibility and efficiency in addressing the complexities of modern front-end design. 

After RTLs are created or generated from HLS tools, designers first use static analysis tools such as Lint~\cite{johnson1977lint} to identify potential errors and then apply various verification techniques, including logic simulation, emulation, and various formal methods (e.g., model checking). These techniques collectively contribute to validating the functionalities of the RTL design faithfully following the design specification. The verification and testing processes, spanning various transformations and stages, are integral components of design flows, typically consuming between 60\% to 70\% of the total engineering efforts allocated. This substantial investment underscores their critical role in ensuring the functionality and reliability of circuit designs. Across diverse abstracts of circuit designs, a plethora of verification techniques are employed, reflecting the nuanced requirements and challenges encountered at each stage of development. For example the C-RTL equivalence checking rigorously compares the RTL implementations against the C-based specification models. This verification method, as evidenced by studies such as \cite{marquez2013formal, mukherjee2017formal}, is frequently applied, particularly in the context of data-path intensive designs, as highlighted by Hector from Synopsis\cite{hector}. Given that circuit designs within this abstract primarily encapsulate hardware behavior while abstracting concrete physical details, theorem provers and SMT solvers emerge as pivotal tools for enhancing verification efficacy~\cite{koelbl2009solver, huang2018instruction}.

Next, the RTL implementation undergoes the next stage in the design flow, wherein logic synthesis tools have revolutionized the way HDL code is transformed into gate-level representations. Logic synthesis typically involves three main steps: elaboration, logic optimization, and technology mapping. The primary objective of logic synthesis is to transform RTL codes into a gate-level netlist that meets specific design constraints while optimizing for power efficiency, maximizing performance, and minimizing the required silicon area, 
all within an acceptable timeframe. 
An indispensable aspect of logic synthesis involves conducting logic and sequential equivalence checks between optimized netlists and their initial counterparts, as underscored by studies such as \cite{mishchenko2006improvements, baumgartner2006scalable, chen2023integrating}. Furthermore, custom equivalence checking techniques have been tailored to cater to specific circuit design requirements, such as those pertaining to clock-gating~\cite{dai2015sequential}.

The collective progression of these front-end design and verification tools has not only streamlined the design process but also expanded the realm of what is possible in digital circuit design. As we navigate increasingly complex design landscapes, these tools have become indispensable in the relentless pursuit of innovation and optimization in digital systems.

\subsection{EDA for Back-End Design}
For modern chip design, the back-end design flow, also referred to as layout design, is depicted in Fig.~\ref{fig:fe_be_flow} b), transitioning from a gate-level or generic technology (GTech) netlist to a finalized layout~\cite{alpert2008handbook}.

This intricate process initiates with technology mapping, where a process library is applied to adapt the synthesized gate-level netlist to a specified technology library, with a keen focus on optimizing PPA constraints. To enhance testability for mass production, testability features such as scan chains, built-in self-test (BIST) circuits, and boundary scan are incorporated into the design. The subsequent phase, physical design, is tasked with establishing the chip's physical layout, entailing floorplanning, power delivery network (PDN) design, placement, clock tree synthesis (CTS), and routing.

\begin{itemize}
    \item \textbf{Floorplanning:} Floorplanning establishes the chip's physical layout by optimizing the placement of major blocks to minimize interconnect lengths and ensure efficient silicon area utilization. It involves strategic arrangement considering timing, power, and thermal constraints to set a foundation for the design.
    
    \item \textbf{Power Delivery Network (PDN) Design:} PDN design ensures stable power supply across the chip, aiming to minimize voltage drop and maintain power integrity. The design of power and ground networks is crucial for delivering power efficiently, with considerations for IR drop, current density, and electromigration.
    
    \item \textbf{Placement:} Placement optimizes the arrangement of standard cells or IP blocks within the floorplan to enhance performance, power, and area. It strategically positions components to reduce wire length, congestion, and considers timing and thermal impacts, employing algorithms to find an optimal configuration.
    
    \item \textbf{Clock Tree Synthesis (CTS):} CTS distributes the clock signal to synchronize the circuit's operations with minimal skew and jitter. Designing a balanced clock distribution network ensures reliable and synchronized performance across the chip.
    
    \item \textbf{Routing:} Routing connects the components based on the established placement and netlist, aiming to complete interconnections without design rule violations or signal integrity issues. It optimizes for shortest paths, minimizes crosstalk and delay, and manages layer assignment and congestion.
\end{itemize}

As chip designs escalate in complexity, the functionalities of back-end EDA tools extend beyond mere layout creation and routing, embracing a multi-faceted optimization challenge. For example, thermal analysis tools empower designers to forecast and address thermal hotspots, guaranteeing the chip's dependable performance across diverse environmental scenarios.  Also, various design for yield (DfY) strategies are required to maximize the manufacturing yield by identifying and mitigating potential yield detractors, performing layout adjustments to address process variations, defect probabilities, and other manufacturing imperfections. Advanced DfY tools and methodologies analyze critical areas, apply lithography-friendly design principles, and optimize the layout to enhance robustness against variations in the fabrication process, ensuring higher yields and reliability of the final product~\cite{li2024ieda}.

Physical verification stands as a critical final step in the back-end design phase, ensuring that the chip layout adheres to all necessary specifications and standards before proceeding to manufacturing. This process involves an array of checks, including design rule checking (DRC), electrical rule checking (ERC), and layout versus schematic (LVS) verification. DRC is essential for validating the layout against a set of predefined rules to ensure manufacturability, focusing on physical dimensions and spacing between circuit elements to prevent fabrication errors. ERC goes a step further by examining the electrical integrity of the design, identifying issues such as signal integrity, power distribution problems, and ensuring the circuit meets its functional requirements. Lastly, LVS verification confirms that the layout accurately reflects the original schematic design, guaranteeing that the physical representation matches the intended circuit behavior. Together, these verification steps identify and rectify potential layout issues, safeguarding the correctness of the final chip.

In summary, the back-end EDA tools have fundamentally transformed the landscape of chip design, empowering designers to craft complex integrated circuits that house billions of transistors operating in unison on a single chip. As semiconductor technology progresses, the significance of EDA tools in the back-end design phase is poised to grow, continuing to fuel innovation and enhance efficiency in chip design research and engineering practices.

\subsection{EDA for Specialized Circuits}

Beyond EDA tools for regular digital circuit designs, the field has witnessed a notable specialization in toolsets designed to meet the unique requirements of standard cells, datapath units, and analog circuits. This evolution underscores the maturation of EDA, providing designers with tailored solutions to optimize these fundamental components efficiently. Specialized EDA tools have become indispensable in addressing the nuanced challenges presented by each component type, enhancing the precision and performance of chip designs.

\begin{figure}[!t]
    \centering
    \includegraphics[width=2.5in]{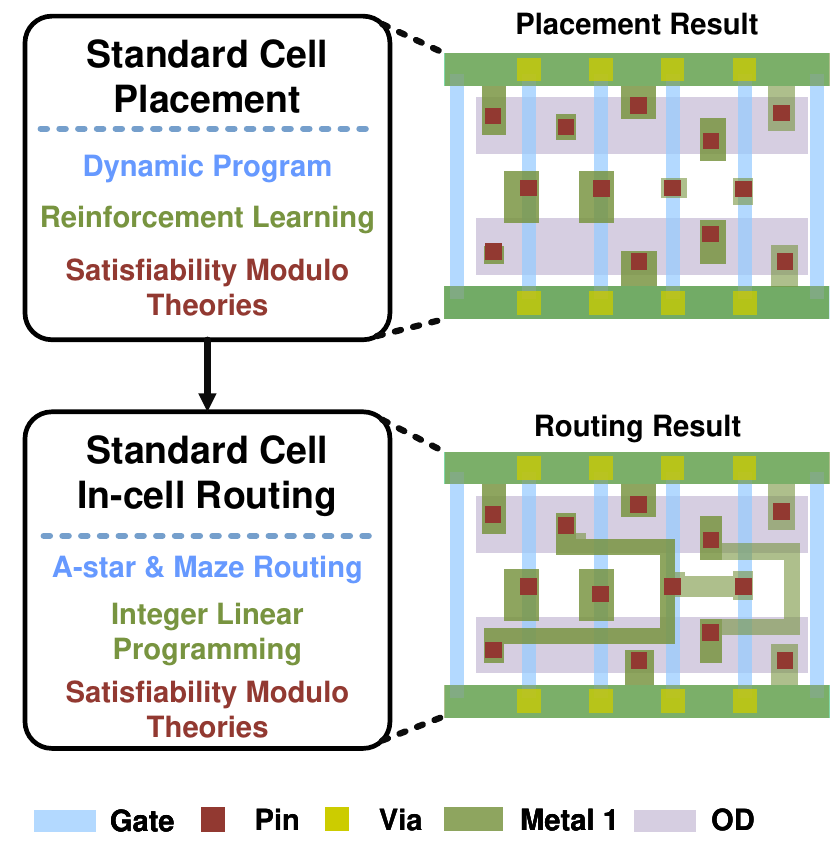}
    \caption{A typical standard cell design flow.}
    \label{fig:stdcell_flow}
\end{figure}

\subsubsection{EDA for Standard Cells}

Standard cells, the building blocks of digital ICs, follow predefined structures that align with a library's specifications, enabling their reuse across diverse designs. The focus of EDA tools in standard cell design is primarily on automating the layout generation process, encompassing crucial steps like placement and in-cell routing, as shown in Fig.~\ref{fig:stdcell_flow}.

The placement process is dedicated to determining the optimal transistor locations within a cell to maximize space utilization while maintaining functionality and performance integrity. The common solution algorithms for the placement includes dynamic program, reinforcement learning, and satisfiability modulo theories. Innovations in placement strategies, as highlighted in~\cite{liNCTUcellDDAAwareCell2019,chengRoutabilitydrivenComplimentaryFETCFET2020}, have introduced methods to expedite this intricate procedure while ensuring routability and design efficiency. In contrast, in-cell routing tackles the intricate task of establishing connections within the cell, a process complicated by the rigorous area constraints of standard cells. The in-cell routing are usually solved by A-star, interger linear programming, and satisfiability modulo theories. This stage demands specialized routing solutions, distinct from those applied to broader digital circuits, to navigate the tight confines of cell layouts. Contributions from~\cite{parkSPSimultaneousPlacement2020,choiPROBE3SystematicFramework2023} have provided targeted approaches to in-cell routing, addressing the unique challenges of standard cell design.

\subsubsection{EDA for Datapath Circuits}

The evolution of datapath circuits, from individual components such as adders, multipliers, multiply-accumulate (MAC) units to the entire datapath, is a testament to the continuous advancements in EDA technologies. Over the years, EDA tools have evolved to address the increasing complexity and performance demands of these critical components.

\textbf{Adders:} Adders serve as the cornerstone of arithmetic operations in digital circuits. The design of adders, from simple ripple-carry to more advanced carry-lookahead and prefix adders, has significantly benefited from EDA tools. These tools employ optimization algorithms to reduce latency, conserve area, and minimize power consumption, crucial for enhancing the overall performance of digital systems. The capability of EDA tools to simulate various adder configurations allows designers to select the most suitable architecture for specific applications, balancing speed with resource utilization.

Specifically, prefix-tree adders, recognized for their efficiency in parallel carry computation, have seen significant development and optimization through EDA solutions. Early designs focused on basic parallel prefix adders like the Kogge-Stone and Brent-Kung adders, which provided a foundation for understanding the balance between speed and area~\cite{beaumont2001parallel}. Recent advancements have introduced more sophisticated designs such as the Sparse Kogge-Stone and Spanning Tree adders, optimizing for both power efficiency and silicon area~\cite{rakesh2019comprehensive}. Datapath compilers have become instrumental in navigating the trade-offs between different prefix-tree configurations, employing algorithmic and heuristic methods to select the optimal structure for a given application scenario.

\textbf{Multipliers:} Multipliers are pivotal in performing fast arithmetic computations, crucial for applications ranging from general computing to specialized tasks in signal processing and machine learning. EDA technologies have facilitated the design of high-performance multipliers by exploring innovative architectures like Booth encoding and Wallace tree multiplication. 

The Wallace Tree technique involves grouping the partial products generated from the multiplication process and then summing these groups in stages, which reduces the overall height of the addition tree and, consequently, the propagation delay. This architecture is particularly favored in digital signal processing (DSP) and graphics processing units (GPUs) where rapid mathematical computations are critical. Over the years, enhancements to the Wallace Tree architecture have aimed at optimizing its layout to minimize area and power consumption while maximizing speed, reflecting the ongoing advancements in EDA tools to meet the evolving demands of semiconductor technology.

\textbf{MAC Units:} The design of MAC units, essential for digital signal processing and deep learning applications, has similarly benefited from the innovations in EDA tools. The integration of optimized adder designs with efficient multipliers within MAC units is critical for achieving high throughput and low latency. EDA tools now utilize analytical models and simulation-based methods to explore various MAC unit architectures, including fused multiply-add (FMA) configurations that perform multiplication and addition in a single operation and pipelined designs, to meet specific performance goals~\cite{jouppi2017datacenter}.

\textbf{Floating-Point Units (FPUs):} Floating-point units are essential for executing arithmetic operations on floating-point numbers, a necessity in applications requiring a wide dynamic range, such as scientific computing, graphics, and machine learning algorithms. 

The evolution of FPUs under the guidance of EDA tools highlights the industry's commitment to addressing the precision, performance, and power efficiency challenges inherent in floating-point operations. Techniques such as pipelining and parallel processing have been integral in enhancing the throughput of FPUs, allowing for simultaneous execution of multiple floating-point operations. Advances in EDA methodologies have facilitated the exploration of novel FPU designs, such as the adoption of FMA units, as in MAC unit designs.

\begin{figure*}[!t]
    \centering
    \includegraphics[width=0.85\linewidth]{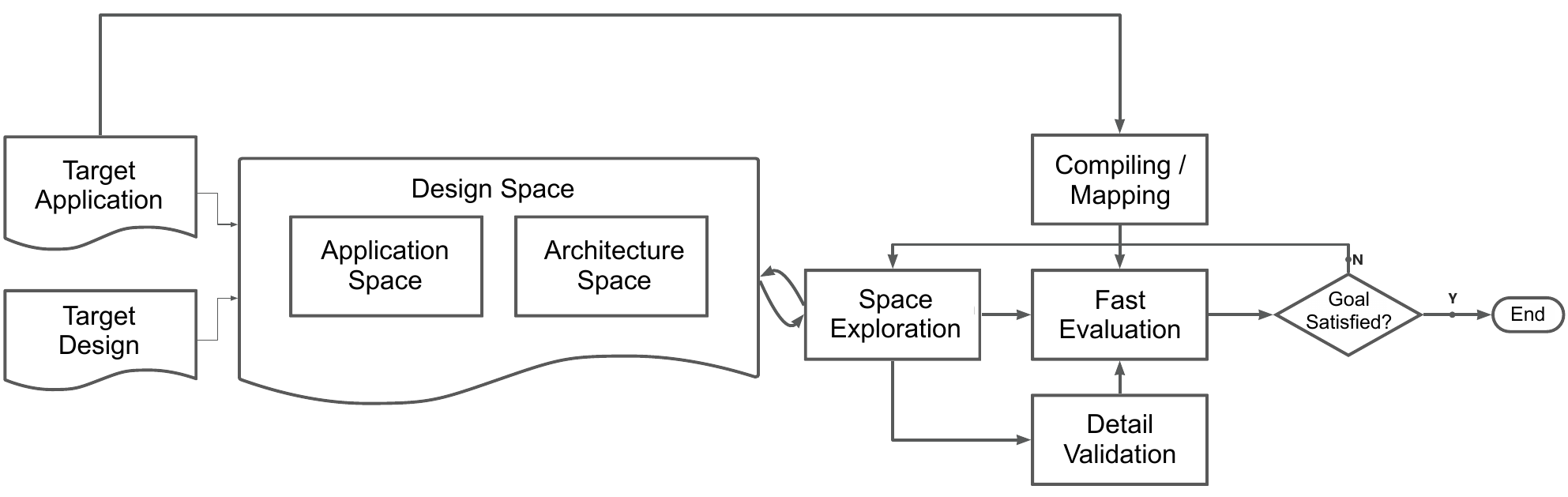}
    \caption{A typical datapath circuits design flow.}
    \label{fig:datapath_flow}
\end{figure*}

\textbf{Datapath circuits:} Beyond individual components, the design of entire datapath circuits, which comprise a combination of adders, multipliers, MAC units, and other logic elements, represents a complex challenge addressed by EDA tools. 

These tools adopt a comprehensive strategy for refining datapath circuits, ensuring seamless integration and peak efficiency among components. As depicted in Fig.~\ref{fig:datapath_flow}, the design journey initiates with pinpointing a target application and its corresponding architectural design, thereby defining a broad and intricate design space. This space might include, for example, CPU tasks like SPEC2017 benchmarks or GPU tasks such as matrix multiplication, targeting either CPU or GPU architectures accordingly. The design space diverges into two principal domains: the application space, outlining application-specific parameters like dataflow patterns or neural network mapping strategies, and the architecture space, detailing the structural and resource parameters, such as CPU issue width or the quantity of MACs in a neural processing unit (NPU).

The intersection of parameters from these domains establishes a ``design point'', which, upon post-compilation application mapping, is subjected to thorough evaluation and validation via cutting-edge EDA tools. This rigorous process iteratively explores and assesses new design points until the exploration goals are achieved.

The progression to new design points is typically steered by optimization algorithms, which have advanced significantly. These optimizations fall into two categories: black box optimization, which proceeds without presuppositions about the design space, often utilizing Bayesian Optimization (BO) for its efficacy in exploring datapath circuit design spaces, and other black box methods like simulated annealing (SA), genetic algorithms (GA), and hill-climbing techniques. Conversely, optimizations that incorporate domain knowledge demand an in-depth understanding of the architecture, aiming for enhancements through precise, targeted adjustments to the datapath. Techniques such as bottleneck analysis have shown to outperform conventional black box approaches by focusing on specific areas for improvement within the datapath architecture.

\subsubsection{EDA for Analog Circuits}

\begin{figure}[!t]
    \centering
    \includegraphics[width=0.9\linewidth]{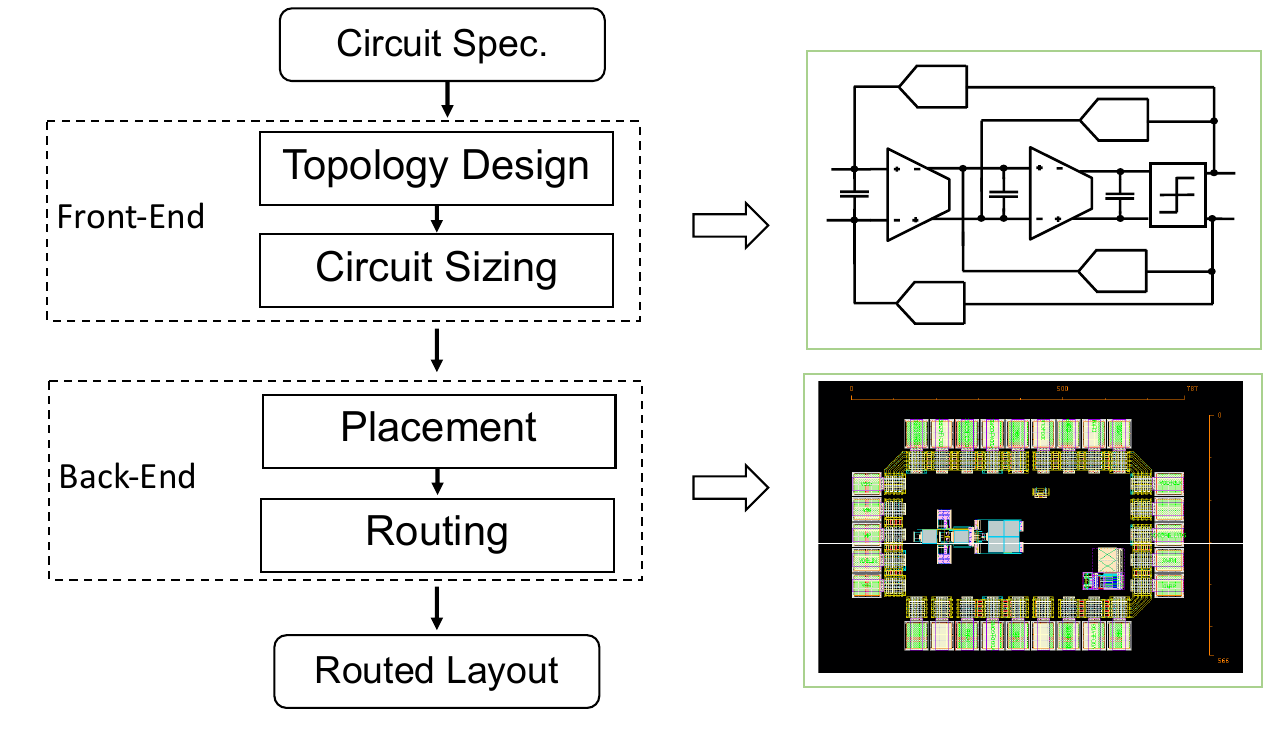}
    \caption{A typical analog IC design flow.}
    \label{fig:analog_flow}
\end{figure}

The design process for analog and mixed-signal ICs significantly differs from digital design, showcasing the unique challenges and complexities of analog circuits.

Fig.~\ref{fig:analog_flow} illustrates a typical analog IC design flow, starting with a detailed set of circuit specifications covering area, power, and performance requirements. The front-end design phase is crucial, establishing the pre-layout circuit netlist that defines the circuit's functionalities via meticulous topology design and device sizing. This phase sets the foundation for the circuit's operational features and optimization criteria.

Moving to the back-end, attention turns to physical layout implementation. Analog physical design, including placement and routing stages similar to digital methods, requires a detailed approach due to analog circuits' sensitivity to parasitics. This phase also incorporates considerations for parasitic effects, component matching, and other layout-dependent factors essential for preserving the circuit's integrity and performance~\cite{Analog_JOS20_Chen}.

A major challenge in analog design is performance optimization, marked by its nonlinearity and the lack of clear functional expressions. Despite these obstacles, the EDA community has significantly advanced the automation of analog IC design over the years. These efforts have covered various areas, such as topology selection or exploration~\cite{Analog_TCAD20_Zhao}, analog sizing~\cite{Analog_TCAS18_Lyu}, analog placement-and-route~\cite{Analog_ICCAD20_Zhu}.

\vspace{5pt}
In summary, the journey of specialized circuit designs encapsulates a dynamic interplay of art and science. As technologies advance and design requirements become more stringent, the role of EDA tools in facilitating efficient, accurate, and innovative design solutions continues to be of paramount importance.

\section{AI for EDA: State-of-the-Art}\label{sec:AI4EDA}

The prowess of deep learning, particularly its capability to discern patterns from historical design data, offers promising enhancements to EDA processes. This modern thrust is propelled by an ambition to harness the extensive repository of design knowledge accumulated across decades to drive superior and more efficient design methodologies.

\subsection{Supervised Learning in EDA}

The utilization of supervised learning in EDA represents a significant stride towards integrating AI into the optimization and estimation of design objectives. This subsection categorizes various supervised AI4EDA solutions based on their application stage within the standard design flow, highlighting seminal works in each category for a focused overview. For those seeking an exhaustive review, references such as~\cite{rapp2021mlcad,ren2022machine} offer comprehensive surveys on the subject.

\subsubsection{Pre-RTL ML Methods}

At the architecture level, supervised ML methods diverge into two primary categories: ML for rapid system modeling and ML as a design methodology.

\begin{itemize}
    \item \textbf{ML for Fast System Modeling:} This approach employs ML to quickly estimate performance and power metrics of circuits and systems. Notable examples include the work by Joseph et al.~\cite{joseph2006construction} and Ithemal~\cite{mendis2019ithemal}, which apply linear and recurrent neural network (RNN) models for CPU performance modeling, respectively. McPAT-Calib~\cite{zhai2021mcpat} enhances CPU power modeling by integrating ML models with the analytical tool McPAT for calibration. PANDA~\cite{zhang2023panda} advances this approach by reducing training data requirements and eliminating dependency on McPAT for power modeling. Boom-Explorer~\cite{bai2021boom} automates design space exploration for the RISC-V BOOM microarchitecture. Beyond CPUs, XAPP~\cite{ardalani2015cross} predicts GPU performance by analyzing dynamic and static properties of single-thread CPU code, while Wu et al.~\cite{wu2015gpgpu} model GPU power by examining kernel scaling behaviors. SVR-NoC~\cite{qian2013svr} focuses on predicting latency and waiting times in mesh-based network-on-chips (NoCs). 
    \item \textbf{ML as a Design Method:} In microarchitecture design, ML techniques facilitate innovative solutions. Shi et al.~\cite{shi2019applying} employ an LSTM model to derive insights from historical program counters for cache replacement using an SVM-based predictor. Pythia~\cite{bera2021pythia} reimagines prefetching as a reinforcement learning challenge, while Hermes~\cite{lu2015reinforcement} leverages ML to predict off-chip load request outcomes. Additional applications include task allocation~\cite{lu2015reinforcement}, power management~\cite{aboughazaleh2007integrated}, and resource management for CPU~\cite{dubach2010predictive} and AI accelerators~\cite{kao2020confuciux}. 
\end{itemize}

In high-level synthesis, the application of ML models for rapidly estimating design metrics has become increasingly prevalent. For instance, Dai et al.~\cite{dai2018fast} focus on timing and resource usage, Pyramid~\cite{makrani2019pyramid} estimates throughput, 
Ustun et al.~\cite{ustun2020accurate} look at operation delay, Zhao et al.~\cite{zhao2019machine} consider routing congestion, and Lin et al.~\cite{lin2022powergear} dedicate their efforts to power consumption analysis.  These studies underscore the versatility of ML in covering a broad spectrum of design metrics, highlighting its capacity to provide comprehensive insights early in the design process.

Moreover, ML's role extends to facilitating design space exploration (DSE) in HLS, exemplified by the work of Ustun et al.~\cite{ustun2020accurate}, Liu et al.~\cite{liu2013learning}, and Meng et al.~\cite{meng2016adaptive}, who implement active learning strategies to navigate the DSE, using predictive ML models as stand-ins for actual synthesis processes. This approach allows for a more efficient evaluation of design alternatives without the need for exhaustive synthesis runs. Additionally, contributions by Kim et al.~\cite{kim2018machine}, Mahapatra et al.~\cite{mahapatra2014machine}, and Wang et al.~\cite{wang2020machine}  demonstrate the integration of ML with traditional optimization algorithms, enhancing their efficacy in navigating complex design spaces.
Sun et al.~\cite{sun2022correlated} introduced a novel approach using correlated multivariate Gaussian process models to capture the intricate interdependencies among multiple objectives across various design fidelities. Yu et al.~\cite{yu2023dse} proposed the IT-DSE framework, leveraging a surrogate model pre-trained on historical design data to refine the search process, illustrating how accumulated design knowledge can be effectively reused to optimize new projects.

In the realm of tensor computations, HASCO~\cite{xiao2021hasco} employs ML for DSE. This methodology optimizes both software programs and hardware accelerators, showcasing ML's capacity to bridge the gap between software and hardware domains to achieve optimized system performance.

\subsubsection{RTL-Stage ML Methods}

At the RTL stage, innovative ML solutions have emerged to predict the PPA without conducting logic synthesis. Initial attempts, such as SNS by Xu et al.~\cite{xu2022sns} and the work by Sengupta et al.~\cite{sengupta}, employ a methodology where the RTL code is converted into an abstract syntax tree (AST) format, from which features are extracted to forecast the design's PPA. Subsequent advancements, including SNS-v2~\cite{xu2023fast} and MasterRTL~\cite{fang2023masterrtl}, claim enhanced accuracy compared to earlier efforts, showcasing the rapid progress in ML applications for RTL analysis. Additionally, there has been a focused effort on applying ML for precise timing or logic estimation~\cite{lopera2022applying,wu2022lostin}.

Power modeling at the RTL stage has also attracted lots of attention. There are two primary categories: design-time power estimation and runtime on-chip power modeling. For design-time estimation, PRIMAL~\cite{zhou2019primal} stands out for offering per-cycle power evaluations tailored to each target design, alongside other notable ML-based approaches~\cite{lee2015dynamic, kumar2019learning}. For runtime power modeling, DEEP~\cite{xie2022deep} introduces an efficient on-chip model that incorporates low-overhead hardware design, utilizing ML to identify power-correlated RTL signals, or `power proxies'. This method, along with other ML-based on-chip power modeling solutions like~\cite{zoni2018powerprobe, pagliari2018all}, demonstrates the potential of ML in creating dynamic power models that adapt to real-time conditions. Moreover, APOLLO~\cite{xie2021apollo} presents a versatile solution applicable to both design-time and runtime scenarios. Simmani~\cite{kim2019simmani} and the early power modeling work~\cite{yang2015early} focus on fast power emulation on FPGA and other platforms, highlighting the broader applicability of ML methods in facilitating efficient power analysis during the design phase.

In the realm of RTL testing and verification, Bayesian networks, as explored by Fine et al.~\cite{fine2003coverage}, offer a probabilistic model-based approach for coverage-based test generation, underscoring the potential of ML in optimizing test planning. Design2Vec~\cite{vasudevan2021learning} advances this further by learning semantic abstractions of RTL designs, facilitating functionality prediction and efficient test generation that notably shortens verification cycles. 
Katz et al.'s~\cite{katz2011learning} decision tree-based method for learning microarchitectural behaviors exemplifies ML's utility in enhancing test stimuli quality.

\subsubsection{Netlist-Stage ML Methods}

Within the netlist stage, supervised learning methods have been leveraged to address a spectrum of challenges including logic synthesis, quality-of-results (QoR) prediction, verification support, and security concerns. 

ML-based models have been particularly effective in assessing synthesis quality and influencing the optimization process. For instance, LSOracle~\cite{neto2019lsoracle} utilizes ML to determine the most appropriate optimizers for various logic networks, thereby enhancing the overall synthesis outcomes. Yu et al.~\cite{yu2018developing} propose to classify and select among multiple random synthesis flows by their quality, subsequently focusing on the most efficacious ones. Their further research~\cite{yu2020decision} extends to evaluating expected delay and area outcomes for synthesis flow candidates, offering a data-driven approach to guide synthesis decisions.

More recent advancements, such as AlphaSyn~\cite{pei2023alphasyn} integrate Monte Carlo tree search with tailored learning strategies for area reduction, showcasing the potential of combining ML with heuristic search techniques for synthesis optimization. Additionally, SLAP~\cite{neto2021slap} targets the enhancement of design timing by identifying and utilizing candidate cuts that lead to improved synthesis results during technology mapping. Their subsequent work~\cite{neto2021read} further demonstrates the ability of ML models to pinpoint post-routing timing critical paths, focusing technology mapping efforts on these areas to minimize delays. DeepGate2~\cite{shi2023deepgate2} develops a pre-trained model that predicts the behavioral correlation of logic gates in netlists and prioritizes SAT-sweeping process to accelerate \textit{fraig} optimization operation.

Following logic synthesis, innovative machine learning solutions are being developed to foresee the post-physical design quality of previously unknown circuit netlists. Tools like Net\textsuperscript{2}~\cite{xie2022preplacement}  pave the way by predicting wirelength and timing information, effectively capturing the implications of placement on the netlist. GRANNITE~\cite{zhang2020grannite} advances this further by facilitating the propagation of RTL toggle rate down to the gate-level netlist, aiming for rapid and accurate average power estimation. Similarly, GRAPSE~\cite{rakesh2023graspe} evaluates average power based on unoptimized and unmapped netlists, showcasing improvements in both speed and precision of power estimation. Recently, DeepSeq~\cite{khan2023deepseq} learns a generic sequential netlist representation that accurately embeds the switching activity behavior and predicts the dynamic power estimation. 

Moreover, ML methods have shown exceptional prowess in deriving high-level abstractions from bit-blasted netlists, unlocking new potentials across various domains within EDA. These high-level abstractions are instrumental in enhancing functional verification, logic minimization, datapath synthesis, and the detection of malicious logic within circuits. For instance, tools like ReIGNN~\cite{chowdhury2021reignn} and GNN-RE~\cite{alrahis2021gnn} utilize ML for reverse engineering tasks, such as identifying state registers and deciphering the functionality of subcircuits. Additionally, ABGNN~\cite{he2021abgnn} leverages graph neural networks to delineate the boundaries of arithmetic blocks in flattened gate-level netlists, while GAMORA~\cite{wu2023gamora}  employs GNNs to infer high-level functional blocks from gate-level data. The success of these methodologies is largely attributed to the capacity of GNNs to discern intricate structural patterns and relationships within netlists, underscoring the transformative impact of ML in enhancing the efficiency and intelligence of EDA processes.

\subsubsection{Layout-Stage ML Methods}

The layout stage presents a crucial phase where ML methods have been increasingly applied to predict or optimize various design metrics such as wirelength, routability, timing, and IR-drop.

\textbf{ML for Placement Stage Enhancements} The placement stage, which determines the optimal locations of macros and standard cells in the layout, is pivotal for achieving desired design metrics. Early applications of ML aimed to augment traditional placement strategies. PADE~\cite{ward2012pade} incorporates support vector machines (SVM) and neural networks for datapath extraction and evaluation, facilitating datapath-aware placement strategies. DREAMPlace, developed by Lin et al.~\cite{lin2019dreamplace}, conceptualizes the placement challenge as akin to training a neural network, thus accelerating the global placement process by harnessing GPU computing capabilities. Building on DREAMPlace, Agnesina et al.~\cite{agnesina2023autodmp} apply multi-objective Bayesian optimization for macro placement design space exploration, demonstrating the potential of ML in enhancing macro-placement outcomes.

ML also assists in predicting design metrics in the later routing phase, benefiting both iterative refinement and early-stage optimization.
Many studies have explored early-stage routability prediction. RouteNet~\cite{xie2018routenet} uses a CNN to forecast the post-routing design rule violations (DRVs), thus avoiding difficult-to-route placements. Another study~\cite{huang2019routability} guides macro placement based on predicted routability. Chang et al.~\cite{chang2021auto} introduce a neural architecture search (NAS) for the autonomous development of routability prediction models, eliminating the need for manually designed machine learning models. Pan et al.~\cite{pan2022towards} propose a federated learning-based approach for routability evaluation, addressing data privacy concerns. To achieve better routability prediction performance, Zheng et al. propose a multimodal neural network Lay-Net~\cite{zheng2023lay}, which aggregates both layout and netlist information.
The ultimate purpose of routability prediction is to assist routability optimization. Liu et al.~\cite{liu2021global} incorporate a fully convolutional network (FCN)-based routability prediction model into the DREAMPlace framework, using it as a penalty factor to explicitly optimize for routability. PROS~\cite{chen2020pros} introduces a routing congestion predictor as a plug-in for commercial placers, effectively adjusting cost parameters to mitigate congestion issues. Moreover, Zheng et al.~\cite{zheng2023mitigating} develop LACO, a look-ahead mechanism designed to address the distribution shift problem in congestion modeling.

Timing is another important metric for placement. 
The field of pre-routing timing prediction at the placement stage has witnessed a range of modeling approaches leveraging various features and machine learning techniques. Studies like those by Barboza et al.~\cite{barboza2019machine} and He et al.
~\cite{he2022accurate} have implemented tree-based methods, incorporating careful manual feature extraction.
TF-Predictor~\cite{cao2022tf} employs Transformers to treat timing paths as sequences, while Guo et al.~\cite{guo2022timing} have devised a customized GNN inspired by static timing analysis mechanisms. Additionally, recent work by Wang et al.~\cite{wang2023restructure} addresses the re-structuring of netlists due to timing optimization, integrating graph data from netlists with layout image information through multimodal fusion.
Moreover, Liang et al.~\cite{liang2020routing} focus on cross-talk prediction, exploring various machine learning models for this purpose.
To reduce turn-around time at the pre-routing stage, Liu~et al.~\cite{liu2023concurrent} propose a concurrent learning-assisted early-stage timing optimization framework called TSteiner, which guides the refinement of Steiner points based on gradients obtained from a GNN-driven timing evaluator.

\textbf{ML for Sign-Off Enhancements:} During the routing and sign-off stages, the precision of sign-off timing, especially using the path-based static timing analysis (PBA), becomes crucial. However, the PBA process is time-consuming, leading to the application of machine learning models for predicting path-based timing based on quicker graph-based analysis (GBA) results. The pioneering work by Kahng et al.~\cite{kahng2018using} was instrumental in predicting PBA from GBA using carefully engineered features and a tree-based model. Subsequent studies, such as ~\cite{cao2022tf,ye2023graph}, have delved into various machine learning models, including transformers and GNN, to enhance the accuracy of GBA-PBA predictions.

Additionally, IR-drop analysis is a critical component in the sign-off stage. Several studies have investigated rapid IR-drop estimation using machine learning, focusing on either static or dynamic analysis to cater to different requirements. For instance, works like IncPIRD~\cite{ho2019incpird} and XGBIR~\cite{pao2020xgbir} concentrate on static IR-drop analysis. In contrast, studies such as~\cite{fang2018machine} target dynamic IR-drop analysis.

\textbf{ML for    Manufacturability Enhancements:} In the field of design for manufacturing (DFM), leveraging ML has become pivotal for bolstering the reliability of lithography and manufacturing processes, with layout patterns often analyzed as images. Studies like GAN-SRAF~\cite{alawieh2020gan}, GAN-OPC~\cite{yang2019gan}, Develset~\cite{chen2023develset}, and L2O-ILT~\cite{zhu2023l2o} use various ML methods to improve mask synthesis printability.
Other works, such as those by Watanabe et al.~\cite{watanabe2017accurate}, Ye et al.~\cite{ye2019lithogan}, Lin et al.~\cite{lin2018data} and Chen et al.~\cite{chen2023physics}, focus on lithography modeling to simulate printed patterns from mask clips.
For identifying layout patterns prone to printing failures like shorts or opens, ML-enhanced lithography hotspot detection is explored in various studies.
For example, Yang et al.~\cite{yang2017imbalance} propose to extract layout features with discrete cosine transform and utilize a
CNN architecture for hotspot detection. The performance is further improved with the proposed bias
learning algorithm because of the imbalanced dataset. Inspired by the object detection problem in computer vision, Chen~et al.~\cite{chen2019lithography} propose to detect multiple
hotspots within large layouts simultaneously. In~\cite{jiang2020efficient}, the binarized neural network
is utilized to speed up the hotspot detection flow. New network architecture is designed based
on residual networks to achieve higher detection accuracy and performance. Additionally, ML further contributes to yield estimation and analysis, as seen in works like Ciccazzo et al.~\cite{ciccazzo2015svm}, Nakata et al.~\cite{nakata2017comprehensive}, and Alawieh et al.~\cite{alawieh2020wafer}.

\subsubsection{Cross-Stage ML Methods}

In addition to stage-specific applications, ML4EDA has significantly impacted the broader task of design flow tuning, garnering substantial interest.

Kwon et al.~\cite{kwon2019learning} introduce a novel approach that blends tensor decomposition with regression analysis to recommend parameters for both logic synthesis and physical design stages, demonstrating ML's capability to streamline design parameterization. FIST~\cite{xie2020fist} utilizes a clustering strategy to automate the adjustment of flow parameters, aiming for enhanced design quality. Furthermore, PTPT~\cite{geng2022ptpt} presents a multi-objective Bayesian optimization framework equipped with a multi-task Gaussian model, significantly improving the design flow tuning process's efficiency.


Verification, a critical component throughout the design process, has also seen the integration of ML to validate circuit design correctness.
Cho et al.~\cite{cho2009eliad} propose an efficient lithography-aware router, which moves lithography verification to the routing stage, effectively enhancing the quality of the printed layout. 

\subsection{Reinforcement Learning in EDA}


Reinforcement learning (RL) in EDA has emerged as a powerful method for navigating the expansive solution spaces inherent in logic synthesis and physical design, often uncovering innovative solutions that surpass traditional, intuition-based approaches. Innovations like Synopsys.ai~\cite{synopsisai} underscore this trend, showcasing AI-driven methodologies that enhance PPA metrics across the design spectrum.



In logic synthesis, Liu et al.'s PIMap framework~\cite{liu2019pimap} exemplifies the application of RL by optimizing LUT-based FPGAs through graph partitioning and iterative synthesis operation selection, leveraging parallelization for efficiency gains. FlowTune, introduced by Yu et al.~\cite{yu2020flowtune}, employs a multi-stage multi-armed bandit (MAB) strategy to constrain the search space and streamline the synthesis process. Pei et al.'s AlphaSyn~\cite{pei2023alphasyn}, utilizing a domain-specific Monte Carlo tree search (MCTS), and Zhu et al.'s approach~\cite{zhu2020exploring}, framing logic synthesis as a Markov decision process (MDP) with a graph convolutional network (GCN), both illustrate the capacity of RL to thoroughly explore synthesis strategies. DRiLLS by Hosny et al.~\cite{hosny2020drills} and subsequent works like those by Peruvemba et al.~\cite{peruvemba2021rl} further extend this exploration, introducing constraints and optimization targets into the RL models to fine-tune synthesis outcomes.
RL has also been applied to logic optimization challenges. For instance, Haaswijk et al.~\cite{haaswijk2018deep} and Timoneda et al.~\cite{timoneda2021late} leverage policy gradient methods and GCNs to optimize majority-inverter graphs (MIGs), showcasing RL's adaptability to various logic structures. 


In physical design, the application of RL ranges from automating chip floorplanning, as demonstrated by Mirhoseini et al.~\cite{mirhoseini2021graph}, to minimizing area and wirelength in floorplanning processes like GoodFloorplan~\cite{xu2021goodfloorplan}. Agnesina et al.'s~\cite{agnesina2020vlsi} use of RL to tune physical design flows for improved PPA metrics and RL-Sizer by Lu et al.~\cite{lu2021rl} for gate sizing highlight RL's potential to refine physical design processes, including timing optimization~\cite{lu2023rl}  and mask optimization in the RL-OPC process~\cite{liang2023rl}. For clock tree synthesis, research efforts are directed toward predicting the quality of the clock network and enhancing timing optimization by leveraging clock skew. GAN-CTS~\cite{lu2019gan} employs a conditional generative adversarial network (GAN) combined with reinforcement learning for predicting and optimizing CTS outcomes.  


\subsection{Leveraging Large Language Models in EDA}




The integration of generative AI, particularly large language models (LLMs), into IC designs is emerging as a transformative trend. By utilizing proprietary datasets, IC design companies can develop AI assistants to enhance and expedite the design process. These tools, capable of providing in-depth insights, automate and refine traditionally manual tasks like design conceptualization and verification. Consequently, a growing body of research explores the application of LLMs in EDA, tackling a broad spectrum of tasks including RTL code generation, task planning, script generation, and bug fixing. While still in the early stages, these studies underscore the profound potential of LLMs to improve the efficiency and efficacy of EDA tools.

This section delves into the use of LLMs for RTL code generation—a key area of focus. 
It categorizes the research into LLM-aided RTL design generation and verification. Additionally, we explore LLM applications in generating EDA scripts and high-level architecture design. 



\subsubsection{RTL Generation through LLMs}

The advent of large language models has ushered in a new era for RTL code generation, offering solutions that have the potential to redefine traditional approaches. 

Early explorations in this domain primarily focused on evaluating models against simple design tasks, hindered by the absence of standardized benchmarks. This challenge has been recently addressed with the introduction of comprehensive benchmarks like RTLLM~\cite{lu2023rtllm} and VerilogEval~\cite{liu2023verilogeval}, facilitating a more robust comparison of LLM capabilities across complex design tasks. RTLLM stands out by providing an open-source benchmark with thirty detailed design tasks, accompanied by ground-truth RTL code for functionality verification. It emphasizes three core objectives: syntax correctness, functional accuracy, and design quality, showcasing a significant leap in performance through innovative prompt engineering techniques like self-planning. Similarly, VerilogEval expands the evaluation framework by gathering Verilog code from diverse sources to construct over 100 test cases. Its approach of collecting additional RTL code for model training demonstrates comparable performance with advanced models like GPT-3.5, yet its training data and model remain unreleased to the public.

Commercial LLMs are utilized for RTL generation, with initial attempts applying GPT-2 for code completion showing promising results~\cite{2021_verilog_completion}. Subsequent developments have introduced tools like ChipGPT~\cite{chang2023chipgpt} and AutoChip~\cite{thakur2023autochip}, which leverage GPT-3.5 to refine code generation through prompt engineering and feedback loops, further reducing the need for human intervention. Chip-Chat's~\cite{Blocklove_2023} achievement in designing a microprocessor with GPT-4 underscores LLMs' potential to autonomously generate hardware description languages.

Recently, the shift towards fine-tuning open-source LLMs presents a viable alternative for customized model development, addressing privacy concerns in VLSI design. Projects like ChipNeMo~\cite{liu2023chipnemo}, RTLCoder~\cite{liu2023rtlcoder}, and BetterV~\cite{pei2024betterv} have demonstrated significant advancements, employing domain adaptation techniques and automated training dataset generation to enhance LLM efficiency and performance for RTL code generation.


\subsubsection{Enhancing Verification with LLMs}

The application of LLMs extends beyond RTL code generation to the verification processes. These models assist in both functional correctness and security analysis, showcasing their versatility and depth in enhancing design validation.

\textbf{Functional Verification through LLMs:}
LLMs have made significant strides in functional verification by translating natural language specifications into SystemVerilog assertions (SVAs). This process ensures that RTL implementations adhere to their intended specifications. Notably, \cite{orenes2023using,sun2023towards} leverage human-written specification sentences alongside RTL designs to generate precise SVAs. AssertLLM~\cite{fang2024assertllm} takes a proactive approach by generating assertions directly from comprehensive specification documents, even before the RTL design phase. This method is complemented by a benchmark set that pairs natural language specifications with golden RTL implementations, offering a robust framework for evaluating assertion generation.
Furthermore, LLMs have achieved success in solving the Boolean Satisfiability (SAT) problem~\cite{zhang2024sola}, which can be applied to verify arithmetic circuits.

\textbf{Security Verification Leveraging LLMs:}
Security validation, critical in identifying and mitigating common vulnerability enumerations (CWEs), has also benefited from LLM integration. Ahmad et al.\cite{ahmad2023fixing} demonstrate the capacity of LLMs to repair hardware security bugs, provided the bug's location is known. 
Further research includes leveraging ChatGPT to recommend secure RTL code\cite{cryptoeprint:2023/212} and employing LLMs in hardware security assertion generation~\cite{kande2023llmassisted}. The latter develops an evaluation framework and benchmark suite that encompasses real-world hardware designs, illustrating LLMs' potential to contribute significantly to security validation efforts.

\subsubsection{EDA Script Generation and Architecture Design}

The versatility of LLM-based solutions in EDA also extends to embrace tasks like EDA script generation and high-level architectural design. 

\textbf{EDA Script Generation:}
ChatEDA~\cite{he2023chateda} introduces an LLM-based agent designed to facilitate EDA tool control using natural language, offering an alternative to traditional TCL scripts. This agent supports a range of operations from RTL code to the graphic data system version II (GDSII), encompassing automated task planning, script generation, and task execution, making EDA tools more accessible and efficient.

\textbf{Architectural Design:}
GPT4AIGChip~\cite{fu2023gpt4aigchip} leverages LLMs to generate C code for AI accelerator high-level synthesis. Similarly, Yan et al.\cite{yan2023viability} examine the use of LLMs in optimizing compute-in-memory (CiM) DNN accelerators, showcasing the model's potential in enhancing computational efficiency. Further extending the scope, Liang et al.\cite{liang2023unleashing} delve into quantum architecture design, exploring the frontiers of quantum computing. SpecLLM~\cite{li2024specllm} contributes to this growing body of work by providing a dataset of architecture specifications at various abstraction levels, investigating LLMs' capabilities in both generating and reviewing these specifications.


\subsection{AI for Specialized Circuits}

The advent of AI4EDA also presents a unique opportunity to redefine the design and optimization of specialized circuits, including standard cells, datapath components, and analog circuits. 

\subsubsection{AI for Standard Cells}

The application of AI in standard cell design, particularly in placement and routing, presents a unique set of challenges due to their high density and strict routability requirements. An AI-assisted approach, utilizing reinforcement learning, has been shown to improve placement sequences and routability, offering better wire length performance~\cite{renInvitedNVCellStandard2021}. Additionally, RL methods have been used to address DRC violations post-routing~\cite{renStandardCellRouting2021}, simplifying the routing process and enabling the use of A-star or maze routing for optimal solutions. Machine learning techniques have also facilitated the adaptation of DRC rules, easing the migration of standard cell layouts across technology nodes~\cite{liangGeneralAutomaticCell2022}. A notable area for AI application is in the evaluation of standard cell layouts, where machine learning models can rapidly assess performance without the need for detailed simulations.


\subsubsection{AI for Datapath Circuits}


Machine learning-based methods are emerging as a powerful tool for optimizing the design of datapath circuits, enabling enhanced efficiency and performance. By leveraging the distinct functionalities and structures of datapath circuits, AI can facilitate a more effective design optimization process.
 


Roy et al.~\cite{roy_learning_2017} employ machine learning to predict the Pareto frontier for adders within the physical design domain. It exemplifies how machine learning can be leveraged for design space exploration, providing insights into optimal design configurations. Utilizing an integrated framework that combines variational graph autoencoders with graph neural processes,~\cite{geng_high-speed_2022-1} develops a novel approach for automatic feature learning of prefix adder structures. This method facilitates sequential optimization, enabling the exploration of Pareto-optimal structures alongside quality metrics. Another study~\cite{cheng_machine-learning-driven_2023} employs multi-perception neural networks to analyze and learn from existing designs and performance data of adders and multipliers. This approach not only achieves high prediction accuracy but also outpaces traditional optimization methods in speed. Moreover, the RL-MUL framework~\cite{zuo2023rl} introduces a novel RL strategy for enhancing multiplier designs. By adopting matrix and tensor representations for the compressor tree and leveraging CNN as the agent, this method allows for dynamic adjustments to the multiplier structure, showcasing the adaptability of AI in complex design optimization.


\subsubsection{AI for Analog Circuits}

AI's integration into analog IC design automation marks a pivotal advancement, enhancing both the efficiency and effectiveness of algorithms. This integration capitalizes on graph and image data representations, mirroring circuit topologies and layouts~\cite{Analog_ICCAD22_Zhu}, to address the challenges inherent to analog design—namely, slow performance evaluation and high search complexity.


\textbf{AI in Analog Topology Generation:} The integration of AI into the generation of analog topologies is revolutionizing the field by speeding up evaluation processes, honing in on more efficient search spaces, and improving optimization techniques. Among the diverse approaches, variational graph autoencoders (VGAEs) have been employed for circuit topologies as showcased by Lu et al.\cite{Analog_DATE22_Lu}, while RL-based methods have been applied to power converters, as demonstrated by Fan et al.\cite{Analog_ICCAD21_Fan}. More broadly, Zhao et al. have utilized RL alongside predefined libraries to address a wider array of problems~\cite{Analog_TCAD23_Zhao}. Poddar et al. have introduced a data-driven strategy for selecting topologies and sizing devices, employing a variational autoencoder (VAE) to synthesize data and thereby reduce simulation expenses~\cite{Analog_DATE24_Poddar}. To tackle the complexities of large circuit design, hierarchical methods are being investigated. Lu et al. have put forward a bi-level Bayesian optimization technique for $\Delta-\Sigma$ modulators~\cite{Analog_TCASII23_Lu}, while Fayazi et al. and Hakhamaneshi have delved into intermediate topology representations and GNN models for voltage node prediction, respectively~\cite{Analog_TCAS23_Fayazi}~\cite{Analog_TCAD23_Hakhamaneshi}. These developments suggest that AI holds significant promise in streamlining the generation of complex topologies, including those of larger circuits comprising multiple sub-circuits.


\textbf{AI in Analog Sizing:} AI is playing a pivotal role in advancing optimization within the realm of analog sizing, notably through the use of ML as surrogate models and RL for direct optimization efforts. ML models, particularly feed-forward neural networks, have been adeptly trained to closely approximate circuit performance metrics. These models, when operated in inference mode, enable the prediction of new, unseen design points, thereby enhancing the efficiency of the search process~\cite{Analog_TCAD21_Budak}. On another front, RL, especially via the GCL-RL algorithm, marries RL techniques with graph neural networks to adeptly optimize analog sizing across varying technological domains. This synergy leverages GNNs' robust capability to encapsulate circuit topologies within the optimization framework~\cite{Analog_DAC20_Wang}. Such methodologies, along with other RL-centric approaches, aim squarely at the intricate balance between global exploration and local exploitation, a balance that is essential for achieving sample efficiency in analog sizing tasks. Innovative strategies, including the use of Voronoi trees for the decomposition of the design space and Monte Carlo tree search (MCTS) for honing in on local search areas, highlight the complex tactics employed to navigate the vast, high-dimensional optimization landscapes with greater efficiency~\cite{Analog_DAC23_Zhao}. The field's progress and the diverse methodologies employed are thoroughly reviewed in a dedicated book chapter, offering a deep dive into the significant advancements and techniques in ML applications for analog sizing~\cite{Analog_Book22_Ahmet}.


\textbf{AI in Analog Layout Automation:} The application of AI in analog layout automation significantly enhances processes such as constraint extraction, placement, and routing, as extensively reviewed in~\cite{Analog_Book22_Burns}.

For constraint extraction in analog layouts, graph-based methodologies are pivotal for identifying symmetry in netlists. These methods encompass graph similarity analysis, edit distance computation, and unsupervised learning for device matching, alongside convolutional graph neural networks for the prediction of layout constraints~\cite{Analog_ICCAD20_Kunal}. A detailed survey on these techniques is provided in~\cite{Analog_ASPDAC22_Zhu_b}.

ML's role in analog layout extends to automating the imitation of expert designs, modeling circuit performance, and optimizing the layout process. GeniusRoute~\cite{Analog_ICCAD19_Zhu} leverages variational autoencoders for making routing predictions that mimic human expertise, impacting various aspects of layout design including well generation~\cite{Analog_DAC19_Xu}, placement strategies~\cite{Analog_DAC21_Gusmao}, and cell generation processes~\cite{Analog_SMACD23_Wang}. CNNs and GNNs are utilized for predicting the performance of designs, thereby optimizing placement and minimizing the dependency on extensive simulations~\cite{Analog_DATE20_Liu, Analog_DATE22_Lin}. The significant impact of ML on performance-driven placement and optimization in analog layout is thoroughly examined in~\cite{Analog_ASPDAC24_Xu}.

Finally, addressing the pre-layout and post-layout simulation gap in analog IC design is vital. ML predicts post-layout parasitics directly from schematics to enhance simulation accuracy and speed up design. For example, ParaGraph~\cite{Analog_DAC20_Ren} employs GNNs for accurate parasitic predictions, using ensemble models for specific value ranges. Early performance assertions using CNNs~\cite{Analog_ICCAD20_Zhang} and layout-aware optimization with BagNet~\cite{Analog_ICCAD19_Hakhamaneshi}, utilizing deep neural networks and evolutionary algorithms, streamline the design process. TAG combines text, self-attention networks, and GNNs for a comprehensive circuit representation, aiding in various predictions~\cite{Analog_ICCAD22_Zhu}.

\section{Large Circuit Models: A New Horizon}\label{sec:LCM}

As discussed in the previous section, AI4EDA solutions have shown remarkable potential, yielding promising outcomes across a spectrum of tasks. However, these solutions predominantly exhibit a task-specific orientation, which, while effective in narrow applications, often limits their scalability and adaptability to the broad spectrum of design challenges.

Venturing into the domain of large circuit models (refer to Fig.~\ref{fig:overview}) marks a bold departure from the previous AI4EDA solutions, moving towards a more integrated and AI-native design process. The term `large' in LCMs signifies both the substantial model size and the vast array of circuit data collected from various EDA stages for circuit pre-training. Such a foundational model concept promises a unified framework that transcends task-oriented limitations, ensuring that LCMs are robust, versatile, and capable of handling the diverse tasks of modern circuit design with limited fine-tuning.

\subsection{Motivation}

The realm of AI4EDA, despite its advancements, faces inherent limitations by primarily repurposing machine learning models from disparate domains to tackle EDA challenges. This approach necessitates the development of distinct models for each specific EDA task. While these models have demonstrated efficacy on benchmark datasets, their ability to generalize to novel designs remains a subject of concern. The unique blend of computation and structure inherent to circuit data requires a nuanced understanding that transcends the capabilities of generic AI solutions. For instance, adapting LLMs for RTL generation without a deep comprehension of circuit design nuances often falls short of achieving optimal PPA results.

The emergence of large foundational models, such as BERT~\cite{devlin-etal-2019-bert}, GPT~\cite{brown2020language}, and MAE~\cite{he2022masked}), has redefined AI's landscape, offering a bifurcated approach of extensive pre-training on diverse data followed by targeted fine-tuning for specific tasks. This methodology has been instrumental in achieving breakthroughs across various data types, heralding a new era of AI applications. The introduction of multimodal foundation models like GPT-4V~\cite{yang2023dawn} and Gemini~\cite{team2023gemini} further exemplifies this trend, facilitating previously unimaginable applications by harmonizing disparate types of data.

Drawing inspiration from these developments, we propose a paradigm shift towards AI-native EDA through the adoption of large circuit models. LCMs, with their focus on learning comprehensive circuit representations, are designed to encapsulate the intricate details and unique characteristics of circuits at every design stage. Echoing the CLIP model's success in bridging text and vision, LCMs aim to forge a similar convergence within EDA, weaving together high-level functional specifications with the minutiae of physical layouts. This holistic approach not only promises to refine the EDA workflow but also aims to significantly reduce time-to-market and enhance the overall design quality such as PPA and circuit reliability.

By championing LCMs, we stand on the cusp of revolutionizing EDA, transcending task-specific limitations, and embracing a future where AI-native solutions drive innovation, efficiency, and excellence in circuit design.

\subsection{Overview of LCMs}

The EDA workflow, extending from initial specification to the detailed final layout, encompasses a variety of circuit design formats, each demanding distinct encoders within the LCMs. These encoders, designed to handle specific modalities – specification, architecture design, high-level algorithms, RTL design, circuit netlists, and physical layouts – are the core components of LCMs. 
To effectively leverage the diverse data inherent to each design modality, LCMs must be pre-trained with a focus on general yet comprehensive design knowledge. This involves not just a superficial understanding but a deep encoding of the nuances present in each modality. For instance, in the circuit netlist modality, the encoded representations must encapsulate both the functional intent and the physical structure of the circuits. This depth of understanding facilitates a more accurate and cohesive foundation for subsequent design tasks. Please refer to Section 5 for details.

The next step in harnessing the power of LCMs involves the fusion and alignment of these unimodal representations to form a cohesive multimodal representation~\cite{baltruvsaitis2018multimodal}. This process is critical in bridging the gaps between disparate stages of the design process, employing advanced techniques such as shared representation spaces, cross-modal pre-training, and innovative fusing strategies. These methodologies aim to synthesize the information captured in individual modalities into a unified, actionable framework that can guide the design process from conception to completion.

Since the specifications, RTL codes, netlists and layout designs are representative formats in front-end and back-end flows, the perspective paper outlines three primary alignment challenges: 

\begin{itemize}
    \item \textbf{Spec-HLS-RTL Representation Alignment:} Utilizing the transformative self-attention mechanism inherent to Transformers, this approach seeks to harmonize the representations of architecture design, high-level C/C++ prototypes, and RTL designs. This unified space enables the coexistence and interaction among these modalities, facilitating a seamless transition across design stages.
\vspace{5pt}
    \item \textbf{RTL-Netlist Representation Alignment:} Inspired by the groundbreaking CLIP model, this challenge leverages contrastive learning and mask-and-prediction training strategies. The goal is to map the embeddings of RTL designs and circuit netlists into a shared latent space, ensuring a coherent progression from logical design to physical implementation.
\vspace{5pt}
    \item \textbf{Netlist-Layout Representation Alignment:} The final alignment challenge focuses on the crucial step of ensuring that the physical layout accurately mirrors the detailed design captured in the netlist. This alignment is vital for the physical realization of the design, embodying the transition from theoretical models to tangible, manufacturable circuits.
\end{itemize}

By confronting these alignment challenges head-on, LCMs promise to revolutionize the EDA workflow, enabling novel applications and methodologies that were previously unattainable. This detailed exploration (please refer to Section 6) sets the stage for a comprehensive discussion on multimodal alignment techniques, further elaborated in subsequent sections, heralding a new era of AI-native circuit design.

\subsection{Opportunities and Potentials}

By accumulating knowledge learned from diverse circuit types and applying cross-stage learning on various design modalities, the potentials of LCMs extend across various aspects of design and verification:

\begin{itemize}
  \item \textbf{Enhanced Verification:} LCMs promise to revolutionize verification by harnessing a deep, cross-stage understanding of circuit designs. This enables more streamlined verification processes, significantly reducing iterations and enhancing the detection of design flaws early in the design cycle. 
  \item \textbf{Early and Precise PPA Estimation:} The comprehensive insights LCMs offer into design data empower them to provide early and accurate PPA predictions. This capability ensures that critical design decisions are informed and strategic from the outset, aligning with optimal design objectives. 
  \item \textbf{Streamlined Optimization:} By pinpointing the true bottlenecks affecting PPA, LCMs can facilitate targeted optimizations. This not only accelerates the design optimization process but also ensures that improvements are effectively implemented across different design stages, enhancing overall design quality. 
  \item \textbf{Innovative Design Space Exploration:} The intelligence imbued within LCMs opens the door to expansive design space exploration. Designers are equipped to discover novel architectures that ingeniously balance PPA trade-offs, fostering creativity and innovation in circuit design. 
  \item \textbf{Generative Design Solutions:}  Perhaps the most revolutionary aspect of LCMs is their potential to underpin generative models capable of autonomously crafting efficient and innovative circuits. This could drastically reduce the time-to-market for new chip designs, offering a competitive edge in the rapidly evolving semiconductor industry. 
\end{itemize}

In essence, LCMs represent not just a technological advancement but a paradigm shift in how circuit design and verification are approached. The full realization of LCMs' potential, however, hinges on the development of sophisticated AI-native techniques for circuit representation learning, challenging the EDA community to explore and harness these untapped capabilities.

\section{Unimodal Circuit Representation Learning}\label{sec:UniLearning}

The journey toward an AI-native EDA paradigm embarks with the essential development of robust unimodal circuit representation learning. These foundational representations are the building blocks for the envisioned multimodal LCMs. This section delves into the nuances of unimodal circuit representation learning, underscoring its indispensable role in establishing a comprehensive and nuanced foundation for sophisticated LCMs. The insights garnered here are paramount for achieving a holistic comprehension of circuit data, which is crucial for the realization of advanced LCMs.

\subsection{Representation Learning for Front-End Design} 
Circuit design commences with the specification and architecture design phase, where the high-level functional intents are formulated. At this juncture, techniques derived from natural language processing are invaluable, transforming specifications into structured, machine-interpretable representations.

As we descend the design hierarchy, representation learning must adeptly adapt to the increasing granularity of detail. At the SystemC and RTL stages, the representation's focus shifts to encompassing the logical and behavioral intricacies of the circuit. In this domain, machine learning paradigms such as LLMs for code, graph neural networks, and hybrid models become instrumental, skillfully capturing the complex logic structures and their interrelations.

\subsubsection{Representation Learning for Architecture Design}

The performance and power consumption of architectures exhibit an intrinsic dependence on specific application contexts. 
In pursuit of optimizing the trade-off among PPA for targeted applications, architectural designers traditionally employ detailed simulation tools complemented by extensive domain-specific expertise. 
This conventional methodology, while comprehensive, tends to be both time-intensive and prone to human errors. 
The advent of LCM presents a novel paradigm, facilitating rapid exploration of architectural design spaces by leveraging insights into the nuanced interactions between application workloads and architectural configurations. 
Thus, it is imperative for LCM to encapsulate application workload representations adaptable to various architectural designs.

Several endeavors have been undertaken in tasks related to architectural design. 
For instance, NPS~\cite{fang2023nps} utilizes a specialized GNN called AssemblyNet for workload representation learning, leveraging both the application's code structure and its runtime states. Trained with a data prefetch task, AssemblyNet identifies the characteristics of typical program slices and minimizes the inaccuracy of sampling-based simulation.
Perfvec~\cite{li2023learning} proposes to learn independent program and architecture representation for generalizable performance modeling. Supervised with an instruction incremental latency prediction task, the yielded model demonstrates applicability on performance modeling across different microarchitectures. On the other hand, several studies have explored the representation of architecture in depth. For instance, GRL-DSE~\cite{yi2023graph} leverages graph representation learning to establish a compact and continuous embedding space for microarchitecture. This approach, utilizing self-supervised learning, enhances the efficiency of identifying optimal microarchitecture parameters. Meanwhile, daBO~\cite{Chirag2023hpca} presents an architecture representation for accelerators enriched with domain-specific knowledge. It involves the manual identification of critical factors that significantly influence the architecture's PPA, and seeks the optimal parameter combinations within this newly defined representation space.

However, existing studies still face challenges in workload and architecture characterization: 
\begin{itemize}
    \item Many models struggle to account for performance-critical factors like branch mispredictions and cache misses, which relate to broader historical states and resist capture through static execution snapshots. An effective LCM must grasp these long-term and complex relationships to accurately represent application workloads.
    \item The intricate relationship between application workloads and power consumption has been underexplored. An ideal LCM would not only integrate power-related factors tailored to varied application workloads, such as flip rates and dynamic voltage fluctuations, but also encapsulate the complex interplay between power consumption and performance, ensuring a cohesive modeling of both aspects.
    \item Current methods primarily concentrate on direct analysis of source code or simulation traces, which overlooks the incorporation of substantial domain knowledge accumulated by experienced architecture designers over the years. A LCM should aim to blend these disparate strands of knowledge, facilitating an enhanced representation learning in terms of accuracy and interpretability.
\end{itemize}

At architectural exploration stage, the focus should be on developing representations that accurately mirror the multidimensional nature of hardware design, capturing not just the static features but also the dynamic interactions within the system. To achieve this, we should employ advanced ML techniques that can process and integrate information from various data sources, including code structure, runtime behavior, and architectural parameters. This process involves constructing multi-layered embeddings that reflect the hierarchical nature of hardware systems, from individual components to the entire architecture. These representations should be learned through a combination of supervised and unsupervised learning tasks, designed to highlight different aspects of the hardware's performance and operational characteristics. By doing so, LCMs can provide a rich, nuanced understanding of the design space, guiding designers towards solutions that optimize performance, power, and area in concert.

\subsubsection{Representation Learning for HLS/RTL}

HLS and RTL represent two pivotal stages in the digital circuit design process. HLS provides a higher-level abstraction, utilizing high-level programming languages such as C, C++, or SystemC to articulate the functionality and behavior of the hardware system. Conversely, RTL offers a more granular view, detailing the data flow between registers and the operations on that data in Verilog or VHDL. Transitioning from HLS to RTL, designers typically employ HLS tools to synthesize the higher-level representation into its detailed RTL counterpart.

To incorporate deep learning in understanding and optimizing HLS/RTL representations, we can explore two innovative methodologies. One method interprets code as a series of tokens, analogous to words in natural language, making it possible to apply NLP techniques to HLS/RTL codes. A particularly effective strategy in this domain is masked language modeling (MLM), where certain tokens are obscured during model training, prompting a Transformer-based encoder (such as BERT) to infer the missing tokens. This self-supervised learning approach yields representations rich in the semantic essence of the hardware design, capturing the functional nuances at both the HLS and RTL levels. Another method may represent HLS/RTL designs as control data flow graphs (CDFGs) offers a graphical perspective, mapping out the control and data dependencies within the design. Here, advanced GNNs come into play, learning from the complex web of interactions and dependencies depicted in the CDFGs. This method allows for the extraction of comprehensive representations that embody the intricate structure and operational logic of the design, providing a solid foundation for subsequent optimization and synthesis tasks.

The former token view is more aligned with the high-level specifications and contains more syntax information. With the application of language models that excel in capturing global relationships, we can get representations that encompass the overall behavior and functionality of the design. Besides, the learned representations will benefit the generalizability and scalability of the attention-based models. On the other hand, the graph view is more aligned with the lower-level gate-level representations and contains more structural and semantic information. Compared to language models, GNNs focus more on extracting local information. 

To enhance the effectiveness of the learned representations, we may consider combining these two views by employing multi-view learning techniques. There are different strategies for integrating these views. The simplest approach involves concatenating the representations obtained from each view and passing them through a multi-layer perceptron (MLP). This allows for the fusion of information from both views, leveraging their individual strengths. Alternatively, a more sophisticated approach is cross-modal prediction, which facilitates deeper interaction between the two views. Through cross-modal prediction, the model is trained to predict one view based on the other view, encouraging the exploration of shared information and dependencies between the representations. By employing multi-view learning techniques, we can maximize the potential of the learned representations and create a more unified and enriched representation of HLS/RTL.

The learned HLS/RTL representations would offer a wide range of applications for various downstream tasks. For instance, they can be leveraged to predict PPA directly from the HLS/RTL, enabling efficient estimation of these crucial design metrics. Additionally, the learned representations can be employed for formal verification to verify the correctness and functional behavior of the design. 


\subsubsection{Representation Learning for Circuit Netlist}

At the netlist level, the design serves as a pivotal junction bridging the front-end design phase with the subsequent back-end processes. Integrating machine learning into logic synthesis, physical design, or verification necessitates a nuanced understanding of the netlist's graph topology alongside gate functionality. This dual focus ensures the netlist encapsulates both the high-level behaviors critical in front-end designs and the intricate structures that profoundly influence PPA in back-end designs.

Initiatives like the DeepGate Family~\cite{li2022deepgate, shi2023deepgate2} stand at the forefront of crafting generalized gate-level representations. The first version~\cite{li2022deepgate} targets circuits in the and-inverter graph (AIG) format and innovatively employs random simulation outcomes to pre-train circuit netlists, with logic-1 probabilities as labels encapsulating crucial functional and structural insights. This pre-training strategy equips DeepGate to capture the core attributes of gate-level circuit designs, allowing for subsequent fine-tuning across a range of front-end applications, such as logic verification~\cite{li2023eda} and design for testability~\cite{shi2022deeptpi}. 

DeepGate2~\cite{shi2023deepgate2} advances this approach by disentangling functional and structural representations within a netlist, learning distinct embeddings for each through specialized labels. Functional embeddings leverage pairwise truth table similarities for supervision, aligning netlists of similar functionalities in close proximity within the functional embedding space. This alignment aids in discerning behavioral similarities and discrepancies. Concurrently, structural embeddings predict pairwise reconvergence, mirroring topological nuances and the complex interconnectivity among logic cells in netlists. Beyond the DeepGate Family, FGNN~\cite{wang2022functionality} introduces a novel contrastive learning task focused on differentiating functionally equivalent from inequivalent circuits, enriching the dataset through strategic perturbations to generate logically equivalent circuit variants.

\begin{figure}[!t]
    \centering
    \includegraphics[width=1.0\linewidth]{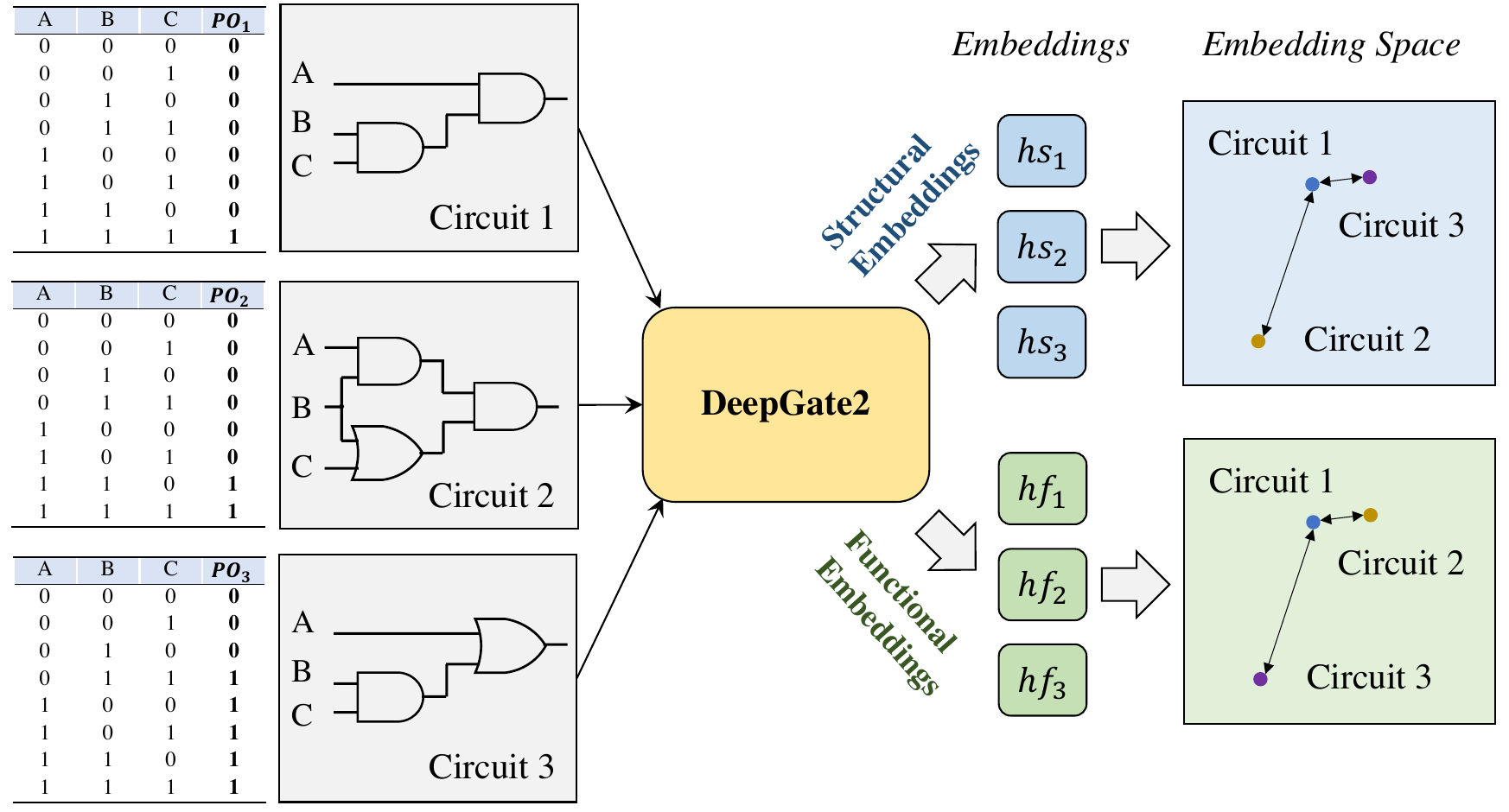}
    \caption{DeepGate2: structural and functional disentangled netlist representation learning. }
    \label{FIG:deepgate}
\end{figure}

After technology mapping, netlists are transformed into a form optimized for the target technology, presenting new challenges and opportunities for representation learning. This stage is critical, as it directly influences the final PPA outcomes of the design.  While we can still formulate the post-mapping netlist as a directed graph and utilize a GNN-based model similar to DeepGate to learn general representations, the complexity of post-mapping netlists, characterized by their technology-specific primitives and configurations, necessitates sophisticated representation learning techniques that can accurately capture the nuances of these transformations. 

While the primary focus of logic synthesis has been on optimizing combinational logic, the sequential behavior of circuits is also a critical facet to represent. DeepSeq~\cite{khan2023deepseq} expands upon the DeepGate technique by elucidating the temporal correlations within sequential netlists. This advancement is facilitated by leveraging both transition and logic-1 probabilities for supervision across each logic gate and memory element, where transition probabilities unveil insights into the circuit's state transition behaviors and logic-1 probabilities illuminate functional and topological characteristics. Such a nuanced approach allows DeepSeq to adeptly encode the complex dynamics and behaviors of sequential circuits, proving instrumental for downstream applications such as netlist-level power estimation and reliability analysis. 

\subsection{Representation Learning for Back-End Design}

Advancing to the physical design stage, representation learning confronts the geometric and spatial intricacies of the circuit layout. Here, convolutional neural networks (CNNs) and vision Transformers (ViT) are particularly adept at capturing the spatial relationships and critical topology in this phase. The objective is to distill the physical design's essence into a representation that not only mirrors the layout's complexities but also yields actionable insights for further optimization and refinement.

The meticulous development of unimodal representations across each design stage knits a rich tapestry of circuit knowledge. Existing studies have explored unimodal learning for prediction of various factors such as routability, IR drop, and lithography hotspots \cite{xie2018routenet, xie2020powernet,yang2019gan}. Although back-end design consists of many design stages with different levels of geometric abstraction, as shown in Fig.~\ref{fig:fe_be_flow} b), existing studies mostly focus on individual stages. There are a few key problems yet to be solved before making back-end representation learning practical in real design applications. For easier understanding, we interpret the layout representation learning task by comparing with computer vision tasks on images. 

Modern layouts consist of rectilinear shapes with a layer property to represent placement and routing information. These shapes need to follow design rules like minimum width, spacing, area, and so on. Detailed of shapes matter. A layout representation encoder needs to capture the detailed changes in layouts. Besides, each shape in a layout is located at a layer. A layer is like an RGB channel of image, so a straightforward way is to encode shapes at each layer into a channel. However, modern layouts often have more than 20 layers, including metal and via layers, which goes far beyond the typical cases of images. 

Unlike images in computer vision which can be resized without losing major information, the dimensions of layouts change with design scales, e.g., $256 \times 256$, $1024 \times 1024$, $4096 \times 4096$, and beyond. Simply resizing a layout like images can lose a lot of information, because individual pixels from layouts of different design scales can correspond to the same geometric resolution defined by manufacturing technologies. A layout representation encoder needs to handle various layout dimensions in a universal way for training on different designs.  

A layout of a chip design contains both geometric and topological (i.e., interconnect) information, its representation needs to align with its circuit graph as well. For instance, if two geometric shapes of adjacent layers (e.g., a metal layer and a via layer) are located at the same positions, they are regarded as connected. A layout representation encoder should be able to identify such topological correlation between shapes. Meanwhile, back-end design has many stages, the geometric information in a layout evolves from abstract to concrete, with more and more details, representations at each stage should align with each other as well. 

These problems raise challenges in learning general representations for back-end design and also call for the multidimensional alignment that is emblematic of LCMs, which will be detailed in the subsequent section.

\section{Harmonizing Representations: A Multimodal Symphony}\label{sec:Multimodality}

In the realm of circuit design, moving away from unimodal representation learning towards a multimodal integration approach offers a fertile ground for innovation. This strategy seeks to merge the distinct representations from each design phase into a cohesive and unified narrative, ensuring a seamless transition across the design stages. Such integration not only maintains a consistent flow of information but also enriches the design process with enhanced coherence.

\subsection{Implementing Multimodal Circuit Alignment}

Central to the concept of multimodal circuit learning is the understanding that all design stages, although distinct in form, share a common functional objective. By applying sophisticated feature extraction and alignment techniques, it becomes possible to overcome the semantic disconnects that typically arise in representation learning. This ensures that the original design intent is not only preserved but also accentuated throughout the entire design lifecycle. The adoption of machine learning models, particularly those leveraging scalable self-attention mechanisms and joint embedding spaces, promises to lead the charge towards a more integrated and holistic approach to circuit design.

A potential solution to achieve this alignment involves the use of masked modeling across different modalities. This technique, inspired by successful applications~\cite{feng2020codebert} in natural language processing, involves selectively hiding parts of the input data across modalities and then training the model to predict these masked portions. By applying this method across circuit design representations—ranging from natural language specifications, high-level algorithms, and RTL implementations to detailed physical layouts can learn a joint representation that captures the essence of the design process at various abstraction levels. This joint representation is crucial for the model to understand the transition from high-level specifications to detailed implementations, enabling it to navigate the complexities of circuit design with greater precision and efficiency.

However, addressing the variability in how a high-level design can be mapped to multiple lower-level implementations, each with PPA characteristics, poses a significant challenge. To tackle this, models need to be equipped with the ability to recognize and evaluate the trade-offs associated with different design choices. Integrating reinforcement learning techniques with the multimodal learning framework can provide a solution. By setting the optimization of PPA metrics as the reward function, the model can learn to navigate the space of possible implementations, identifying solutions that best meet the specified criteria. Furthermore, incorporating attention mechanisms can enhance the model's ability to focus on relevant features across modalities, thereby improving its capacity to predict implementations that not only meet functional requirements but also optimize for PPA objectives. Through these methods, the implementation of multimodal alignment in circuit design can become not just a theoretical concept but a practical tool for advancing the field.


Considering the vast differences between design modalities, aligning them in a single step is a formidable challenge. To address this, we propose a phased approach to multimodal alignment, partitioning the process into three distinct phases: ``Spec-HLS-RTL Representation Alignment," ``RTL-Netlist Representation Alignment," and ``Netlist-Layout Representation Alignment." Since the RTL design contains high-level semantics and netlist is more suitable for aligning with the following backend designs, this strategy employs these two designs as intermediaries, facilitating a more manageable and focused alignment process. By breaking down the alignment into these stages, we can concentrate on specific transitions within the design flow, allowing for a more tailored application of machine learning techniques to each phase. This phased approach not only makes the task of alignment more feasible but also ensures that each stage of the design process is optimally aligned, leading to more coherent and efficient design outcomes. Through careful implementation of this strategy, we aim to bridge the gap between the various design modalities, ultimately fostering a more integrated and seamless circuit design environment.

\subsection{Spec-HLS-RTL Representation Alignment}

The transition from conceptual specifications to RTL implementation involves a complex journey through natural language specifications, architectural exploration, high-level languages such as SystemC, and hardware description languages like Verilog and VHDL. Leveraging LCMs within a multimodal framework would significantly refine this transformation across different stages, boosting the quality, efficiency, and pace of the design process. LCMs orchestrate a unified representation space that ensures the harmonious integration of front-end design elements across various formats. This unified approach not only streamlines the capture of intricate relationships among circuit components but also accelerates design generation, enhances optimization efforts, and streamlines verification, embodying a leap forward in circuit design methodology.

One of the paramount applications of aligning representations at this stage is the potential substantial improvement in RTL generation. As discussed earlier, existing RTL generation techniques merely fine-tune large language models on HDL code, a process that lacks circuit-specific understanding. With the aligned representations, we could devise a more sophisticated tokenization strategy for HDL code, paving the way for a deeper understanding and representation of hardware design intricacies. This method transcends the capabilities of existing approaches by generating RTL code that is not only syntactically accurate but also semantically rich, closely aligned with the initial specifications and high-level design intentions. Such advancements promise to elevate the precision and applicability of automatically generated RTL, ensuring designs are both optimized and verifiable from the outset.

Furthermore, the C2RTL verification process benefits significantly from the aligned representation facilitated by LCMs, addressing a pivotal challenge in the transformation from high-level specifications to RTL. This verification phase necessitates a thorough comparison of functional behaviors across natural language specifications, high-level programming languages like C/C++, and RTL implementations. Traditionally, within the EDA framework, this comparison has been both labor-intensive and prone to errors, largely due to the disconnect between the abstract, functional descriptions at the high level and the detailed, hardware-specific implementations at the low level. Bridging this gap between high-level and low-level circuit representations has been a long-standing challenge for the EDA community.

The adoption of LCMs with multimodal alignment into this process introduces a transformative approach to C2RTL verification. By harmonizing the representations of the circuit's functionality across different stages, these models significantly streamline the verification process. LCMs can identify and resolve discrepancies by meticulously comparing the generated RTL representation against its high-level counterparts. This capability is enhanced by the transformer technology, renowned for its ability to attend selectively to various parts of the input based on their relevance. Such focused attention allows the models to concentrate on areas where discrepancies between the intended functionality and its RTL implementation are most pronounced, offering precise insights and resolutions to designers. This method not only reduces the time and effort traditionally associated with C2RTL verification but also increases the accuracy and reliability of the verification process, marking a significant advancement in ensuring circuit design integrity and performance~\cite{kande2023llmassisted, orenes2023rtl}.

\subsection{RTL-Netlist Representation Alignment} 

The RTL-Netlist Representation Alignment stage is crucial for bridging the gap between RTL, AIG netlists, and post-mapping netlists. This alignment paves the way for numerous applications, significantly impacting early PPA estimation, design optimization, and verification processes.

One of the primary benefits of RTL-Netlist alignment is the enhancement of early PPA estimation. By aligning representations from the RTL design phase through to the netlist level, designers can gain insights into the potential power, performance, and area characteristics of their designs much earlier in the development cycle. This early insight allows for more informed decision-making, enabling adjustments to the design that can lead to optimal PPA outcomes. Such proactive adjustments can significantly reduce the need for time-consuming and costly revisions at later stages, streamlining the design process and accelerating time-to-market.

Beyond early PPA estimation, RTL-Netlist alignment also opens the door to more sophisticated design optimization strategies. By having a clear view of how RTL designs translate into netlist implementations, designers can identify and address inefficiencies at a much deeper level. This insight enables the application of targeted optimizations that can improve the overall quality and efficiency of the design. Moreover, leveraging machine learning models trained on aligned data sets allows for the automation of some optimization tasks, further enhancing the design efficiency and effectiveness.

Finally, the alignment between RTL and netlist representations significantly benefits the verification process. With a comprehensive understanding of how design intentions are manifested in the netlist, verification teams can develop more accurate and efficient testing strategies. This alignment ensures that the verification process is not only faster but also more thorough, reducing the likelihood of errors slipping through to later stages. The ability to detect and address potential issues early on, based on a deep understanding of the aligned representations, is invaluable in maintaining design integrity and reliability.

\subsection{Netlist-Layout Representation Alignment}


The aspiration to align the circuit netlist with its physical layout is not merely an ambition but a transformative step in EDA.
In traditional EDA workflows, the netlist, which represents the logical abstraction of a circuit, and the physical layout, which represents the concrete geometries of the circuit, have been treated as separate entities.
However, the increasing complexity of modern integrated circuits have highlighted the need for a tighter integration between the logical and physical domains.

By aligning the netlist with the physical layout, designers can gain a deeper understanding of the relationship between the logical function and physical form of a circuit. 
This alignment enables a unified perspective of the design, where the logical and physical aspects are considered together, rather than in isolation.
It allows designers to analyze and optimize the design from a holistic standpoint, taking into account the impact of physical constraints on logical functionality, and vice versa. Another key benefit of achieving this alignment is the ability to revolutionize the verification process.
Traditionally, verification has been performed separately for the logical and physical domains, leading to potential mismatches and design errors.
With a multimodal approach that considers both domains simultaneously, designers can detect and resolve issues that arise due to the interaction between the logical and physical aspects of the design.
This comprehensive view of the design across stages ensures that the final product meets the desired specifications and performs as expected.

Furthermore, the integration of logical and physical information opens up new possibilities for design optimization. 
By presenting an integrated picture of the entire design space, designers can explore a wider range of possibilities and make more informed decisions.
This comprehensive perspective allows designers to identify and address potential bottlenecks or issues early in the design process, leading to improved quality and efficiency.
A specific example of the significance of integrating netlist-layout information is in pre-routing timing prediction.
Pre-routing timing prediction aims to accurately evaluate potential sign-off timing violations in the early stages of the design process, reducing design cycles and avoiding costly iterations. Traditionally, pre-routing timing evaluation methods, such as static timing analysis, have primarily focused on netlist information, which represents the interconnections between cells in a design.
However, these methods often overlook the crucial role that layout information plays in timing prediction.
As most timing optimization techniques require space to insert or resize gates, the circuit layout that reflects spatial information has a large impact on sigh-off timing performance.
Neglecting layout information can lead to inaccurate timing predictions and sub-optimal design decisions.
Through netlist-layout representation alignment, LCM can provide more accurate estimates of sign-off timing performance.
This enables designers to identify and address timing issues early in the design process, reducing the likelihood of sign-off violations and the need for time-consuming iterations.

\vspace{5pt}

In summary, the evolution towards a multimodal symphony in circuit design represents not just a technical advancement, but a reimagining of how design processes can be optimized for efficiency, innovation, and coherence. The potential for such an approach to revolutionize the field lies in its ability to harmonize disparate data types and design stages into a single, unified framework, paving the way for breakthroughs in design methodology and implementation.

\section{Pioneering LCM Applications} \label{sec:Cases}

\begin{table*}[!t]
\centering
\caption{Solving time comparison between Ours and~\cite{een2007applying} on LEC cases.} \label{TAB:Comp_LEC}
\renewcommand\tabcolsep{2.0pt}
\footnotesize
\begin{tabular}{@{}l|lll|llllll|llllllll@{}}
\toprule
\multicolumn{1}{c|}{\multirow{2}{*}
{Case}} & \multicolumn{3}{c|}{Baseline}     & \multicolumn{6}{c|}{~\cite{een2007applying}}                                                    & \multicolumn{8}{c}{Ours}                                                                  \\
\multicolumn{1}{c|}{}                      & \# Vars   & \# Clas   & $\mathcal{T}_{solve}$     & \# Vars   & \# Clauses & $\mathcal{T}_{trans}$  & $\mathcal{T}_{solve}$   & $\mathcal{T}_{all}$     & \textbf{Red.}    & \# Vars   & \# Clas   & $\mathcal{T}_{agent}$   & $\mathcal{T}_{trans}$  & $\mathcal{T}_{solve}$   & $\mathcal{T}_{all}$     & \textbf{Red.}  & \textbf{Red.*}  \\ \midrule

I1      & 42,069  & 105,711  & 322.46   & 5,616    & 54,529   & 5.31     & 51.49   & 56.80  & 82.39\%            & 3,160    & 31,281   & 9.27    & 5.62     & 4.43    & 19.26 & 94.03\%            & 66.08\%  \\
I2      & 44,949  & 112,954  & 708.97   & 6,052    & 60,573   & 5.61     & 147.85  & 153.46 & 78.35\%            & 4,112    & 41,873   & 9.81    & 6.12     & 4.41    & 20.81 & 97.07\%            & 86.44\%  \\
I3      & 42,038  & 105,629  & 531.94   & 5,612    & 54,825   & 5.21     & 109.89  & 115.10 & 78.36\%            & 3,849    & 37,329   & 8.37    & 5.61     & 2.91    & 17.56 & 96.70\%            & 84.74\%  \\
I4      & 37,275  & 93,678   & 289.89   & 5,038    & 49,805   & 4.61     & 90.05   & 94.66  & 67.35\%            & 3,478    & 34,013   & 7.32    & 5.11     & 2.50    & 15.01 & 94.82\%            & 84.14\%  \\
I5      & 30,087  & 75,537   & 172.79   & 4,006    & 38,069   & 3.91     & 38.77   & 42.67  & 75.30\%            & 2,311    & 22,473   & 4.78    & 4.31     & 1.10    & 10.50 & 93.92\%            & 75.39\%  \\ \midrule
\textbf{Avg.}                              & \textbf{} & \textbf{} & \textbf{405.21} & \textbf{} & \textbf{}  & \textbf{} & \textbf{} & \textbf{92.54} & \textbf{77.16\%} & \textbf{} & \textbf{} & \textbf{} & \textbf{} & \textbf{} & \textbf{16.63} & \textbf{95.90\%} & \textbf{82.03\%} \\ \bottomrule
\end{tabular}
\end{table*}

While extensive empirical data are yet to be available, the potential applications of LCMs can be vividly illustrated through hypothetical scenarios and conceptual frameworks. The narrative examples presented in this section serve to bridge the gap between abstract concepts and tangible applications, offering a glimpse into the transformative impact LCMs could have on the EDA field. 

\subsection{Circuit Learning for SAT} 

The Boolean Satisfiability (SAT) problem identifies if there exists at least one assignment that makes a given Boolean formula to be \textit{True}. SAT problem acts as a fundamental problem in many areas, especially in the EDA fields, such as logic equivalence checking, model checking, and testing. Over the past few decades, the SAT community has advocated adopting the conjunctive normal form (CNF) as the \emph{de facto} standard format for problem instances and developed numerous advanced CNF-based SAT solvers~\cite{sorensson2005minisat, fleury2020cadical}. However, the efficacy of CNF-based solvers recently encounter bottlenecks in solving hard SAT problems, prompting past research to explore circuit-based solvers or strategies as a potential breakthrough. In this section, we aim to demonstrate the impact of the large circuit model on SAT solving. 

First, the circuit netlist serves as a natural representation of SAT problems within the field of EDA and also can be efficiently derived from various combinatorial optimization problems. Inspired by an early endeavor~\cite{een2007applying}, a circuit-based universally efficient reformulation mechanism could significantly reduce the complexity before solving these problems. The LCMs, especially the uni-modal netlist encoders, are capable of capturing the structural features across various netlist distributions. Exploiting this knowledge allows for the exploration of a global transformation flow based on reinforcement learning, ultimately minimizing the overall complexity of the solving process. 

Table~\ref{TAB:Comp_LEC} shows our preliminary results when applying the netlist encoder to accelerate SAT solving for industrial logic equivalence checking cases I1\--I5. In the Baseline setting, the instances are solved directly using the Kissat solver~\cite{fleury2020cadical}. We denote the RL agent runtime, transformation time, and solving time as $\mathcal{T}_{agent}$, $\mathcal{T}_{trans}$ and $\mathcal{T}_{solve}$, respectively, measured in \textit{seconds}. The overall runtime, which sums up all three components, is denoted as $\mathcal{T}_{all}$ in \textit{seconds}. Additionally, we list the number of variables (\# Vars) and clauses (\# Clas), the reduction in $\mathcal{T}_{all}$ compared to Baseline (\textbf{Red.}) and compared to~\cite{een2007applying} (\textbf{Red.*}). 
The solving time is reduced by $96.14\%$ and $82.03\%$ on average, respectively.

Second, gate-level embeddings proficiently encapsulate the logical correlations among gates within a circuit netlist, ensuring that gates sharing functional similarities are closely aligned within the embedding space. This alignment allows for a precise representation of logical connections between variables in the SAT formulation. By integrating these gate-level embeddings, we can highlight and utilize the discerned correlations to expedite the SAT-solving process. This is achieved by embedding these correlations as additional constraints in the initial SAT problem instances, thereby enhancing the solver's efficiency.

Third, traditional heuristic strategies (e.g., branching heuristics) predominantly depend on the correlation between variables in CNF representations, which cannot preserve the circuit's topological structure. Recent advancements, such as~\cite{shi2023deepgate2}, showcase the effectiveness of a unimodal netlist encoder in capturing the intricate gate-level logic correlations within circuit netlists.

Building upon the above, the LCMs excel in identifying gate-level functional relationships within circuit netlists based on the unimodal netlist encoders. 
By harnessing the power of LCMs, new and efficient circuit-based SAT-solving strategies can be developed, ultimately improving the overall performance and effectiveness of heuristic designs. 

\subsection{LCM for Logic Synthesis} 

Logic synthesis stands at the crossroads of multiple representations and sophisticated algorithms, such as truth tables, sum-of-products, binary decision diagrams (BDDs), and directed acyclic graphs (DAGs), with none asserting complete dominance. This diversity underscores a fundamental challenge: selecting and optimizing the most effective representation for a logic function. Herein lies the transformative potential of LCM. By learning and internally representing the same logic function across diverse formats, LCMs exhibit unparalleled adaptability. Their deep understanding of intricate relationships and optimization pathways within logic synthesis allows for a flexible approach to representing complex logic functions. This adaptability becomes instrumental in handling multifaceted inputs and expressions of logic, showcasing LCMs' capability to revolutionize the representation and optimization of logic functions in a way previously unattainable.

As we venture into the realm of nanometer-scale technologies, the significance of technology-independent optimizations becomes increasingly pronounced, focusing on metrics like the number of literals and logic depth in DAGs for area and delay evaluation. Marrying these optimization strategies with the physical realities of the technology landscape introduces a new layer of complexity. LCMs are poised to address this challenge head-on by predicting physical characteristics such as timing, area, and power more accurately. Integrating physical awareness, LCMs offer a groundbreaking tool in logic optimization, enabling designers to base their decisions on a nuanced understanding of circuit behavior. This foresight not only refines optimization strategies but also facilitates superior PPA trade-offs, marking a leap forward in logic synthesis.

The crux of technology mapping, especially in FPGA and ASIC design, lies in navigating the constraints of heterogeneous logic blocks, interconnect resources, and optimal cell selection while balancing the PPA trade-offs. Addressing structural bias during technology mapping demands meticulous algorithmic strategies. LCMs herald a new era in technology mapping by leveraging their ability to learn from diverse logic representations and adapt mapping strategies to the nuanced requirements of the input data and technological constraints. Their versatility in overcoming structural biases through contextually aware mappings, coupled with the iterative feedback loop and physical information integration, offers tailored insights for refining mapping strategies. LCMs' scalability further underscores their effectiveness in managing complex circuit designs, presenting a compelling case for overcoming longstanding challenges in technology mapping.

In essence, the conceptual application of LCMs in logic synthesis promises a shift towards more efficient, accurate, and adaptable design processes, positioning them as the cornerstone for the next generation of circuit design methodologies.

\subsection{LCM for Equivalence Checking} 

Equivalence checking stands as a critical verification step in digital circuit design, ensuring that functionality is preserved through synthesis or manual modifications. Traditional methods, while reliable, struggle with scalability in the face of increasingly dense designs and the complex optimizations required to meet PPA goals. Here, LCMs emerge as a transformative solution, offering a paradigm shift towards interactive equivalence checking that enhances the efficiency and effectiveness of the process.

LCMs have the unique potential to revolutionize this domain by enabling an end-to-end interactive equivalence checking process. This approach is particularly beneficial for ECO optimizations and custom design styles, where the goal extends beyond functional equivalence to include high-quality design modifications. Leveraging their deep understanding of circuit semantics, LCMs can offer insightful recommendations for design adjustments and patches during the interactive ECO phase. Drawing from extensive training on diverse circuit data, LCMs can identify underlying patterns and rules of successful designs, suggesting targeted modifications to resolve detected discrepancies. These suggestions are not only based on historical success but are also ranked according to their anticipated impact on PPA, empowering designers with informed choices that align with their specific objectives.

Furthermore, the iterative nature of LCMs means that these recommendations can be refined based on designer feedback, creating an efficient feedback loop that streamlines the equivalence checking and modification process. This iterative engagement not only accelerates the identification of viable design solutions but also enhances the overall quality of the final design.

In addition to transforming equivalence checking into an interactive dialogue, LCMs hold promise for augmenting existing equivalence checking systems. Traditional algorithms have exploited the empirical distribution of circuit designs, wherein current practices include: 1) partitioning and selecting fine-grained proof strategies~\cite{cadence-lec}, 2) adapting various encodings from a problem instance to a canonical solver instance~\cite{zou2024beswac}, and 3) employing design-specific equivalence checking strategies (e.g., for multipliers~\cite{chen2023datapath-cec}). These solutions remain limited by the need for hand-crafted heuristics and specialized strategies. For instance. LCMs, with their ability to automatically understand design intent and manage the distribution of design data, can act as a neural backbone for these systems. They can manage various heuristics in formal solvers or function as a neural scheduler for task distribution, significantly enhancing the performance and efficiency of equivalence checking processes.

This dual approach—transforming equivalence checking into an interactive process and augmenting existing systems—highlights the pioneering potential of LCMs. By leveraging the power of LCMs, designers can navigate the complexities of modern circuit verification with greater ease and precision, promising to elevate the verification process to new heights of efficiency and effectiveness.

\subsection{LCM for Physical Design} 

Physical design is the stage that converts the logical representations of a circuit into the physical representations. In this stage, a physical layout is generated by partitioning, floorplanning, placement, and routing. This process requires solving many NP-hard combinatorial optimization problems and is extremely complex and time-consuming.
As the scale of an electronic design keeps increasing and the feature size keeps shrinking, traditional approaches to physical design face serious challenges.
LCMs, on the other hand, could provide new perspectives on processing the physical representations of an electronic design and even new methodologies in dealing with these tricky combinatorial optimization problems.

A trained LCM could offer guidance to placement and routing for wirelength, routability, and timing optimization. 
Consider the placement optimization process, where traditional methods have leveraged unimodal information for guidance, from gradient prediction~\cite{liu2022xplace} to routing congestion forecasting~\cite{xie2018routenet}. Common practices involve transforming layout features into image-like data for machine learning model predictions, often employing vision-based models like CNNs and Vision Transformers. Yet, this approach may overlook crucial interconnect information, given the challenges vision-based methods face in preserving topological details alongside spatial relationships.
Recent explorations into multimodal representations for physical design, however, illuminate a promising path forward. Studies like LHNN~\cite{ROUTE_DAC2022_Wang} introduce dual GNNs to capture both topological (circuit interconnections) and spatial relationships, merging these insights in latent space. Similarly, Lay-Net~\cite{zheng2023lay} proposes substituting the GNN with CNN for spatial analysis, capitalizing on the superior spatial awareness of vision-based methods. Despite these advancements, LCMs have the capacity to move beyond merely integrating topological and spatial relationships. By aligning with additional modalities, designers gain the unprecedented ability to pinpoint layout hotspots at earlier design stages and implement preemptive countermeasures. This proactive approach facilitated by LCMs allows for the early identification of potential issues related to timing, power, and thermal management, enabling adjustments before they escalate into more significant challenges.  

To sum up, LCMs could learn the underlying characteristics of a physical representation and reveal new directions for design and optimization.
More excitingly, they have the potential to serve as the foundation of new learning-based heuristics and revolutionize the traditional way of physical design, eliminating the burden of constantly designing new algorithms.

\section{Tailoring LCMs for Specialized Circuits}\label{sec:Special}

Exploring specialized circuit domains reveals a diverse array of unique designs that extend beyond the standard digital circuits typically encountered in EDA workflows. Standard cell designs, datapath circuits, memory macros, and analog circuits possess distinct characteristics that necessitate custom approaches. The expansion of LCMs into these specialized arenas heralds a promising enhancement for design efficiency and optimization.

\subsection{Large Circuit Models for Standard Cells}
Standard cells form the fundamental building blocks of digital designs, comprising basic logic gates and complex combinational functions. Their design is critical for the overall performance and power efficiency of the chip. LCMs in this domain could leverage generative models to propose new cell architectures that optimize for a variety of constraints, including power, performance, area, and even novel objectives like robustness to process variations. Furthermore, these models could predict the impact of cell design changes on the higher levels of the design hierarchy, enabling a holistic approach to optimization.

For the front-end design of standard cells, LCM can be employed for library pruning and cell characterization. The requirements for standard cell libraries differ between high-performance circuit design and low-power circuit design~\cite{library_prune1,library_prune2}. Historically, designers have often relied on experience and extensive simulations to select a subset for a new cell library. LCM can leverage existing selection experiences to better choose suitable cells for the specific design scenario. Additionally, it can leverage generative models to continuously explore new topological architectures, subsequently refining the generation process based on SPICE simulation results as feedback for continual improvement. Characterization is the most time-consuming step in standard cell design, requiring extensive SPICE simulations to generate liberty libraries. However, the significance varies across different PVT corners and standard cells. Therefore, accuracy-aware supervised learning can enhance the overall precision of libraries while reducing runtime by prioritizing the importance of different corners and cells~\cite{aging_lib_ml,solido_characterization_suite}.

\subsection{Large Circuit Models for Datapath Circuits}


Datapath circuits, essential for performing arithmetic and logical operations within microarchitectures, stand at the core of performance-critical computing. These components notably benefit from bit-level optimization, necessitating a detailed focus on timing and power constraints. 

In datapath design, the optimization of a myriad of parameters is crucial for balancing PPA effectively. 
Acting as surrogate models during design space exploration, LCMs can significantly streamline the optimization process. LCMs specifically tailored for datapath circuits offer a promising approach by employing specialized architectures adept at understanding the complexities of arithmetic operations. This enables them to enhance logical efficiency while optimizing the physical layout. Through training on diverse datasets, encompassing both synthetic and real-world datapath designs, LCMs pave the way for exploring innovative datapath configurations that extend beyond traditional design methodologies.

Specifically, LCMs bring a new dimension to datapath design evaluation, offering a comprehensive and accurate analysis that transcends traditional limitations. Traditional evaluation methods often rely on a constrained set of benchmarks, limiting the scope of assessment. In contrast, LCMs draw upon a broad and profound understanding of target applications, facilitating a more extensive evaluation of datapath designs. This broad perspective enables a ``shift-left" in the evaluation process, providing early and insightful assessments that encompass not only architectural considerations but also subsequent stages like placement and routing. Such a shift enhances the overall efficiency and effectiveness of the evaluation process.

Moreover, LCMs' deep domain knowledge in both the logical functions and physical implementations of datapath circuits allows them to automatically pinpoint optimization bottlenecks. This capability not only accelerates the design optimization cycle but also furnishes designers with critical insights for further enhancements. By integrating LCMs into the datapath design process, engineers can achieve a level of optimization and efficiency previously unattainable, heralding a new era in the evolution of datapath circuits and their implementation in modern microarchitectures.

In summary, LCMs' detailed grasp of datapath complexities allows them to offer strategic recommendations that go beyond parameter adjustments, influencing the architectural framework of the circuit's RTL design. The ultimate goal is to utilize LCMs for the automated generation of circuit datapaths, tailored to specific Process Design Kits (PDKs) and targeted software applications, thereby revolutionizing the design process.

\begin{figure}[!t]
    \centering
    \includegraphics[width=\linewidth]{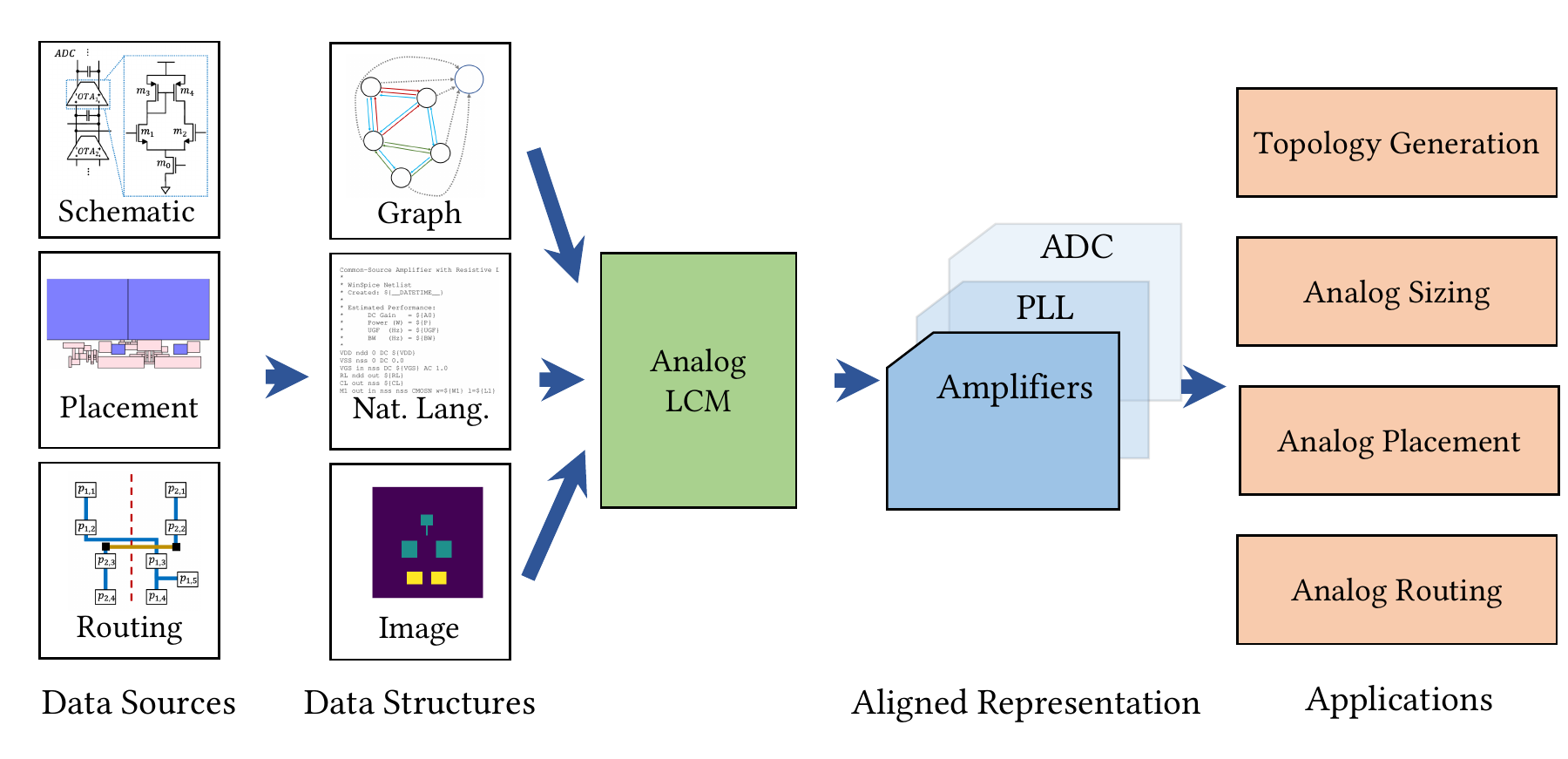}
    \caption{Overview of large circuit models for analog EDA.}
    \vspace{-5pt}
    \label{fig:analog_lcm}
\end{figure}

\subsection{Large Circuit Models for Analog Circuits}
Analog EDA shares similarities and differences with digital EDA. Like digital workflows, analog EDA encompasses front-end netlist design and back-end layout design. Analog LCMs also demand holistic solutions that span different design flow stages. Conversely, analog circuits exhibit distinct data structures and performance evaluations compared to their digital counterparts, which are primarily logic-driven. In analog circuits, device-level topology and physical implementation are crucial. The sub-structure of transistors, capacitors, and resistors determines circuit functionality, making the detailed graph structures (such as network motifs) and device parameters essential for capture by analog LCMs. Moreover, analog circuits involve various types and evaluations, with different performance evaluations requiring specific circuit implementations.

Analog circuit design is an art that melds intricate knowledge of device physics with the subtleties of the intended application. LCMs for analog circuits must capture this depth of knowledge, translating it into models that can navigate the analog design space with its continuous variables and stringent performance metrics. These models could predict analog behavior from device-level up to system-level specifications, assist in layout generation, and automate the tedious tuning process of analog parameters. By doing so, LCMs could drastically reduce the design time and enhance the performance of analog circuits, which remain a bottleneck in mixed-signal chip design.

TAG~\cite{Analog_ICCAD22_Zhu} represents an early effort to develop a circuit representation model for analog EDA. It introduces a netlist embedding mechanism and a ``pretrain-then-finetune" strategy to apply embedding vectors across various applications. However, it lacks a unified, aligned representation across all design stages, with its effectiveness constrained by the initial pretraining target, layout distance. Its potential applications are somewhat limited compared to a comprehensive LCM for analog circuits.

Fig.~\ref{fig:analog_lcm} presents our vision for future analog large circuit models. These models are inherently multimodal, capable of processing various data structures from different design stages. Text and graph structures can represent a netlist, while images may be used for layout designs. An Analog LCM converts these inputs into vectors, mapping the designs to a unified embedding space. The generated circuit embedding vectors can then support various downstream tasks across different circuit types (such as amplifiers, PLLs, and ADCs), catering to a range of applications from topology design to routing.

\section{Challenges and Opportunities: The Dual Edges}\label{sec:Challenges}

Embarking on the quest for LCMs unveils a realm filled with both challenges and opportunities. The journey is strewn with hurdles like data scarcity, scalability issues, and interoperability with existing EDA tools, yet each challenge surmounted paves the way for uncharted opportunities.

\subsection{Data Issues}
Data scarcity stands out as a critical hurdle, given the dependency of LCMs on extensive, high-quality datasets for training. The realm of circuit design, particularly at the granularity required for effective LCM operation, suffers from a lack of publicly available data, posing risks of overfitting and undermining the models' generalization capabilities. Tackling this issue head-on, we introduce three possible solutions. 

First, innovative data augmentation techniques emerge as a key solution. For instance, equivalent circuits can be generated through systematic circuit transformations, effectively expanding the dataset without the need for additional real-world data. This approach not only enhances the diversity of training material but also deepens the model's understanding of circuit variability and design principles.

Second, on the synthetic data front, leveraging LLMs to generate realistic RTL code presents an exciting opportunity. This strategy involves using LLMs' advanced generative capabilities to create new RTL designs, which can then serve both as additional training data and as benchmarks for further refining the RTL generative models themselves. This creates a self-reinforcing loop where LCMs continually improve through iterative training on both real and synthetically generated data. Such a mechanism not only addresses the issue of data scarcity but also contributes to the evolution of more sophisticated and capable generative models, marking a significant leap towards fully realizing the transformative potential of LCMs in the EDA landscape.

Third, the development of community-driven platforms for data sharing and collaboration could significantly alleviate the scarcity issue. By fostering an ecosystem where academia and industry share circuit data and design challenges, the field can collectively advance the state of LCM research, ensuring a diverse and comprehensive dataset that mirrors the multifaceted nature of electronic design automation.

In essence, while data scarcity presents a formidable challenge, it also opens the door to a range of inventive strategies that not only address the immediate issue but also enrich the EDA domain. Through collaborative efforts, technological advancements, and a commitment to innovation, the potential of LCMs in revolutionizing circuit design remains within reach.

\subsection{Scalability and Interoperability}

Scalability emerges as a pivotal challenge in the realm of LCMs, especially as we delve into complex, vast-scale circuit designs that define the next generation of electronic devices. The quest for scalability is not just about accommodating larger designs but also about enhancing computational efficiency and sophistication in model architecture. This involves pioneering hierarchical modeling techniques that can intuitively decompose complex designs into manageable submodules, algorithmic optimizations that streamline model training and inference, and the implementation of parallel processing strategies to distribute computational workload effectively. Each of these advancements contributes to a robust foundation, equipping LCMs to tackle increasingly ambitious design projects while maintaining precision and efficiency.

Moreover, as LCMs grow in complexity and capability, ensuring their interoperability with the existing mosaic of EDA tools becomes paramount. The modern circuit design ecosystem is a tapestry of specialized design flows, tools, scripts, libraries, and technologies, each contributing to various stages of the design process. Bridging the gap between the innovative potential of LCMs and the established practices of current EDA workflows necessitates a concerted effort for deeper collaboration between the AI research community and EDA professionals. Such collaboration aims to weave AI-driven methodologies seamlessly into the fabric of EDA, enhancing tool compatibility, data exchange protocols, and user interfaces. This symbiotic relationship stands to not only streamline the integration of LCMs into existing design pipelines but also to catalyze the mutual evolution of both AI technologies and EDA tools and methodologies, heralding a new era of design automation that is both more intelligent and intuitive.

\subsection{New Opportunities}

Beyond merely enhancing existing EDA tools, LCMs present the exciting prospect of birthing entirely new categories of EDA tools, ones that could fundamentally alter how design, verification, and optimization are approached. 

One of the most promising opportunities presented by LCMs is the ability to conduct early-stage, precise PPA estimation. Traditionally, accurate PPA metrics could only be determined after substantial design progress, often at the post-synthesis or post-layout stages. LCMs, however, can predict these critical metrics much earlier in the design process, leveraging aligned representations among modalities. This capability allows for more informed decision-making at the outset of a project, guiding design choices in alignment with PPA objectives and significantly accelerating the optimization cycle. Early-stage PPA estimation not only enhances design efficiency but also enables a more agile response to evolving design requirements and constraints.

LCMs also enable a paradigm shift towards cross-stage verification, a holistic approach that transcends the conventional, compartmentalized verification processes. Traditional EDA methodologies often treat verification as a stage-specific task, siloed within the design flow. However, LCMs, with their comprehensive understanding of circuit knowledge across various stages, facilitate a unified verification framework. This cross-stage verification can detect inconsistencies and errors early in the design process, reducing the iterative cycles typically required to rectify such issues. By leveraging the predictive power of LCMs, designers can ensure coherence and fidelity from the initial specifications to the final physical layouts, significantly streamlining the verification process.

Moreover, LCMs unlock the potential for generative design, particularly for well-structured circuits such as datapath units. Datapath units, with their regular structures and predictable performance metrics, are ideal candidates for LCM-driven generative design approaches. LCMs can generate optimal circuit configurations that meet specified criteria, exploring a vast design space that might be infeasible for human designers to cover comprehensively. This generative capability can lead to innovative circuit designs that optimize PPA metrics, potentially discovering novel architectural solutions that traditional design methodologies might overlook. Furthermore, generative design facilitated by LCMs can automate aspects of the design process for these structured circuits, reducing manual effort and enabling a focus on higher-level design challenges.

Finally, the synergy between large language models and LCMs presents a particularly promising area of exploration. LLMs, with their advanced natural language processing capabilities, can serve as intuitive, conversational interfaces for designers, translating high-level design specifications into actionable insights and suggestions; While the LCMs, with their deep understanding of circuitry and design principles, can analyze and optimize the granular details of the netlist, ensuring that the final design aligns with the desired performance, power, and area constraints. This collaborative interaction between LLMs and LCMs allows for a seamless transition from abstract design concepts to concrete, optimized circuit representations. Bridging the gap between high-level design intent and detailed technical execution, this synergy enables a more holistic and integrated approach to circuit design. 

In summary, the development of LCMs is fraught with challenges, yet each obstacle surmounted brings the EDA community one step closer to realizing the full potential of these innovative models. The promise of LCMs to significantly streamline the design process, elevate design quality, and accelerate the development of cutting-edge electronic systems highlights the critical importance of addressing these challenges. 
\section{Conclusion}\label{sec:Future}

As we navigate the evolving landscape of AI-driven EDA, the potential of large circuit models emerges as a beacon of innovation, promising to redefine the paradigms of circuit design and analysis. 

Specifically, we advocate for a paradigm shift from task-oriented AI4EDA methodologies to more integrated, AI-native foundation models. LCMs stand at the crossroads of this transition, offering a holistic representation that encapsulates the multifaceted aspects of circuit design—from logical structuring to physical realization. The promise of LCMs lies in their ability to harness deep learning for capturing the intricate dependencies and characteristics of large-scale circuit netlists, thereby facilitating more efficient, accurate, and innovative design strategies.

Looking ahead, the journey toward fully realizing the potential of LCMs is laden with a vast array of research problems waiting to be addressed. From the refinement of representation learning techniques to accommodate the unique circuit characteristics at various design stages, to the development of scalable, effective alignment models capable of interpreting and optimizing complex netlists, the field is ripe for exploration. 

In conclusion, the dawn of AI-native EDA heralded by LCMs presents a transformative vision for the future of circuit design and analysis. By embracing this new frontier, we stand to unlock unprecedented levels of efficiency, creativity, and precision in the creation of the next generation of electronic devices.


%



{
\balance
\bibliographystyle{IEEEtran}
\bibliography{./Top.bib, ./ref.bib, ./analog_ref.bib, ./stdcell_ref.bib}
}

\end{document}